\documentstyle[twocolumn,epsfig,fleqn,amstex,rotating]{mn}
\newcommand{\hi}{\mbox{H\ {\footnotesize I}}}

\newcommand{\hei}{\mbox{He\ {\footnotesize I}}}
\newcommand{\heii}{\mbox{He\ {\footnotesize II}}}
\newcommand{\heiii}{\mbox{He\ {\footnotesize III}}}

\newcommand{\nhi}{\mbox{\scriptsize H\ {\tiny I}}}

\newcommand{\nhe}{\mbox{\scriptsize He}}

\newcommand{\nheii}{\mbox{\scriptsize He\ {\tiny II}}}

\newcommand{\Ang}{\ \mbox{\AA}}

\def\etal{{et al.\ }}

\def\lsim{~\rlap{\raise 0.4ex\hbox{$<$}}{\lower 0.7ex\hbox{$\sim$}}~}
\def\gsim{~\rlap{\raise 0.4ex\hbox{$>$}}{\lower 0.7ex\hbox{$\sim$}}~}
\def\dd{{\rm d}}
\def\fnq{f_{\dot{N}_Q}}
\def\lz{L_\ast(z)}
\def\l0{L_\ast(0)}
\def\sz{S_\ast(z)}
\def\s0{S_\ast(0)}
\def\ezz{e^{\zeta z}}
\def\exz{e^{\xi z}}
\def\exs{e^{\xi z_\ast}}
\def\omg0{\Omega_0}
\def\omgl{\Omega_\Lambda}
\def\omgb{\Omega_b}

\def\a2{\alpha^{(2)}}
\def\ahi{\alpha_{\nhi}^{(2)}}

\def\aheii{\alpha_{\nheii}^{(2)}}
\def\trho{ {T-\rho}}

\def\fclmp{f_{\rm clump}}

\def\lcdm{\Lambda\rm{CDM}}

\def\lph{l_{\rm ph}}
\def\nq{\dot{N}_{\rm Q}}
\def\tq{t_{\rm Q}}
\def\xj{x_{\rm J}}

\def\ne{n_{\rm e}}
\def\nabs{N_{\rm abs}}
\def\ntrs{N_{\rm trs}}

\def\mpc3{\ {\rm Mpc^{-3}}}
\def\gpc3{\ {\rm Gpc^{-3}}}

\def\erghzs{\ {\rm erg\ Hz^{-1}s^{-1}}}

\def\smpc{\ {\rm s^{-1}Mpc^{-3}}}
\def\hmpc{\ {\rm h^{-1}Mpc}}

\def\phts{\ {\rm phot\ s^{-1}}}

\def\ev{\ {\rm eV}}

\begin{document}

\title[]
{Patchy He\ {\LARGE II} reionization and the physical state of the IGM}
\author[Gleser {\it et al.}]{Liron Gleser$^{1,2}$, Adi Nusser$^{1,3}$, Andrew J. Benson$^4$, Hiroshi Ohno$^5$, \& Naoshi Sugiyama$^5$\\\\
$^1$Physics Department, Technion, Haifa 32000, Israel\\
$^2$The Institute of Astronomy, Madingley Rd, Cambridge CB3 OHA, UK\\
$^3$The Institute for Advanced Study, Princeton, NJ 08540\\
$^4$University of Oxford, Astrophysics, Keble Road, Oxford OX1 3RH, UK\\
$^5$Division of Theoretical Astrophysics, National Astronomical Observatory Japan, Mitaka, 181-8588, Japan}
\maketitle

\begin{abstract}
We present a Monte-Carlo model of \heii\ reionization by QSOs and its
effect on the thermal state of the clumpy intergalactic medium
(IGM). The model assumes that patchy reionization develops as a result
of the discrete distribution of QSOs. It includes various recipes for
the propagation of the ionizing photons, and treats photo-heating
self-consistently. The model predicts the fraction of \heiii, the mean
temperature in the IGM, and the \heii\ mean optical depth --- all as a
function of redshift. It also predicts the evolution of the local
temperature versus density relation during reionization. Our findings
are as follows: The fraction of \heiii\ increases gradually until it
becomes close to unity at $z\sim 2.8-3.0$. The \heii\ mean optical
depth decreases from $\tau\sim 10$ at $z\gsim 3.5$ to $\tau\lsim 0.5$
at $z\lsim 2.5$. The mean temperature rises gradually between $z\sim
4$ and $z\sim 3$ and declines slowly at lower redshifts. The model
predicts a flattening of the temperature-density relation with
significant increase in the scatter during reionization at $z\sim
3$. Towards the end of reionization the scatter is reduced and a tight
relation is re-established. This scatter should be incorporated in the
analysis of the Ly$\alpha$ forest at $z\lsim 3$. Comparison with
observational results of the optical depth and the mean temperature at
moderate redshifts constrains several key physical parameters.
\end{abstract}



\section {Introduction}
\label{sec:introduction}

The physical properties and the thermal history of the
intergalactic-medium (IGM), play an important role in shaping the
observed structure in the universe. A proper modelling of the
evolution of the IGM is therefore essential for the interpretation of
the rapidly accumulating data on the high redshift universe. These
data include the secondary temperature and polarization fluctuations
in high resolution measurements of the cosmic microwave background
(CMB), and absorption features in QSO spectra, which probe the ionized
and neutral gaseous components, respectively. Since the distribution
of gas follows the underlying mass, these data can be used to
constrain the linear mass power spectrum (Croft \etal 1998, 2002;
McDonald \etal 2000, 2004; Viel, Haehnelt \& Springel 2004).

The gaseous component of the IGM follows, by large, the distribution
of the underlying mass distribution of the gravitationally dominant
dark matter (DM). The temperature, ionization state and other physical
properties are, however, determined by energetic feedback incurred by
QSO and galactic activities. This feedback is manifested in
photo-ionization and heating of the IGM by radiation emitted by QSOs
and galaxies, and in mechanical energy from supernovae explosions of
massive stars produce galactic winds that can shock-heat the IGM and
enrich it with metals (e.g. Cowie \& Songaila 1998; Aracil \etal
2004).

Absorption features in QSO spectra indicate that mechanical energy in
galactic winds stirs up the IGM only in the vicinity of
galaxies. Observations (Adelberger \etal 2003) have shown that winds
can evacuate the neutral hydrogen from within a distance of up to
1-2$\hmpc$ away from Lyman break galaxies (LBGs). Away from galaxies,
analysis of the Ly$\alpha$ forest strongly suggests that the moderate
density IGM which makes up most of the volume in space, is only
affected by ionizing UV photons emitted by QSOs and galaxies.

The gaseous IGM is mainly made of a primordial composition of hydrogen
and helium with the former being the dominant element. At $z\lsim
1000$ these elements recombined and remained neutral until the
appearance of stars (and possibly very high redshift QSOs) able to
produce sufficient photons to cause significant
reionization. Measurement of the polarization anisotropies of the CMB
by the Wilkinson Microwave Anisotropy Probe (WMAP) satellite call for
an ionized IGM at redshifts as high as $z\sim 17\pm 5$ (Kogut \etal
2003; Spergel \etal 2003). The Gunn-Peterson trough in QSO spectra
seems to occur at $z\approx 6$ (Becker \etal 2001; Djorgovski \etal
2001) and might imply a complex reionization history of the IGM with
more than a single phase of reionization (Cen 2003; Wyithe \& Loeb
2003b). Ionizing radiation from galaxies is also sufficient to singly
ionize helium (Benson \etal 2002) at $z>6$. In any case observations
of the Ly$\alpha$ forest show that by $z\sim 6$ most of the hydrogen
in the universe was photo-ionized and that the IGM is in a quiet state
governed by the gravitational drag of the dark matter and the
photo-heating of the ionizing radiation (see Rauch 1998 for a
review). During this stage a tight relation between the temperature
and density is established (e.g. Theuns \etal 1998). The enhancement
of QSO activities peaking at $z\sim 2$ is expected to disturb this
quiet state by the production of hard photons energetic enough to
doubly ionize helium (e.g. Zheng \etal 2004). Modelling double helium
reionization is the focus of the current paper. One of our basic
assumptions is that QSOs are the main sources for \heii\ reionization
at low redshifts. This assumption is sustained by the evidence from
observations pointing to \heii\ reionization at redshifts concurrent
with the escalation of QSO activities. Although star formation
activity also peaks at similar redshifts, radiation from galaxies is
too soft to contribute significantly to \heii\ reionization (Benson
\etal 2002, Wyithe \& Loeb 2003a), unless Population III stars are
considered (e.g.  Venkatesan, Tumlinson \& Shull 2003).

In recent years commendable efforts have been made to observe an
\heii\ Gunn-Peterson trough, at the Ly$\alpha$ wavelength of 304{\AA},
in quasar absorption spectra. Estimation of the \heii\ optical depth
from the spectrum of the quasars Q0302-003 ($z=3.286$, Dobrzycki \etal
1991; Heap \etal 2000) and of HE 2347-4342 ($z=2.885$, Smette \etal
2002; Zheng \etal 2004) indicate a sharp decline in the optical depth
at redshift $z\sim 2.8$. At $z\gsim2.8$ the optical depth $\tau\gsim
4$ and at $z\lsim2.7$ the optical depth $\tau\lsim 1$. This decline
implies that \heii\ reionization occurred at $z\sim 3$. The optical
depth of the quasar HS 1700+64 ($z=2.743$, Davidsen \etal 1996)
$\tau=1.00\pm 0.07$ at redshift $z=2.4$, agrees with the general
picture described above. It is important to stress that there may be
other explanations for the large fluctuations in the \heii\ optical
depth. For example, in Smette \etal (2002), a model is described where
optical depth fluctuations along the line of sight are a local effect
due to nearby ``soft'' and ``hard'' sources, and are unrelated to the
global reionization of the IGM.

We aim at a detailed modelling of the state of the IGM during \heii\
reionization. Two approaches can be adopted in achieving this goal.
The first is to employ hydrodynamical simulations that include
radiative transfer and a good recipe for modelling the distribution of
QSOs. An effort towards this goal has been made by Sokasian \etal
(2001, 2002) who implemented their radiative transfer numerical code
in a cosmological simulation of a box of $67\hmpc$ (run by
V. Springel). The simulations offer a valuable insight into the
evolution of ionized regions and their topology. The effect of \heii\
reionization on the thermal properties of the IGM have not been
modelled in these simulations. This is because radiative transfer is
not merged self-consistently with the gas dynamics in the
simulations. This is a technical limitation which is expected to be
overcome in the near future. However, the simulations are limited by
the available CPU power, preventing a thorough exploration of the key
physical parameters. The alternative approach is to develop fast
semi-analytic models that capture the essential ingredients of the
reionization process. This approach has been adopted to study a
variety of other problems in structure formation. (e.g. Bi \& Davidsen
1997; Kauffmann \etal 1997; Somerville \& Primack 1999; Kauffman \etal
1999; Cole \etal 2000; Benson \etal 2001). The basis of semi-analytic
models is the synthesis of the knowledge acquired from diverse
simulations and analytic reasoning. This synthesis allows us to study
a given physical problem with greater detail than individual
simulations offer directly. In this paper we propose a Monte-Carlo
model for studying \heii\ reionization assuming QSOs are the sole
sources of the ionizing radiation. For an assumed cosmology and a form
for the QSO luminosity function, the model provides detailed
information about the physical state of the IGM. The model is easy to
run and allows a thorough exploration of the key physical process. We
confront the model with available observational data, and provide
predictions for the equation of state of the IGM, i.e., the local
temperature-density relation (hereafter $\trho$ relation). We compare
the \heii\ mean optical depth, $\tau(z)$, to the optical depths
calculated from the \heii\ Ly$\alpha$ forest of the three quasars
mentioned above (Q0302-003, HE 2347-4342 and HS 1700+64) and use this
to constrain our model. We also compare the mean temperature, $T(z)$,
from our model to the temperatures at the mean density calculated from
nine quasars' Ly$\alpha$ spectra (Schaye \etal 2000) to place further
constraints.

The paper is organized as follows. A short theoretical background is
presented in \S~\ref{sec:overview}. In \S~\ref{sec:model} we describe
our model which we use to investigate the process of patchy \heii\
reionization and its effect on the thermal evolution of the IGM. In
\S~\ref{sec:compare} we try to fit observational measurements of the
mean temperature and the mean optical depth of the IGM with our model
for the power-law $\lcdm$ cosmological model implied by the WMAP data
(Spergel \etal 2003). We also compute the local temperature-density
relation for different redshifts. In \S~\ref{sec:cosmmodels} we repeat
the same computation and comparison for a $\lcdm$ model with running
spectral index (RSI) (Spergel \etal 2003) and for an open CDM (OCDM)
cosmological model. We conclude with a summary and a discussion of the
results in \S~\ref{sec:discussion}.


\section{An overview of He\ {\small \bf II} reionization}
\label{sec:overview}

\begin{figure}
\centering
\mbox{\epsfig{figure=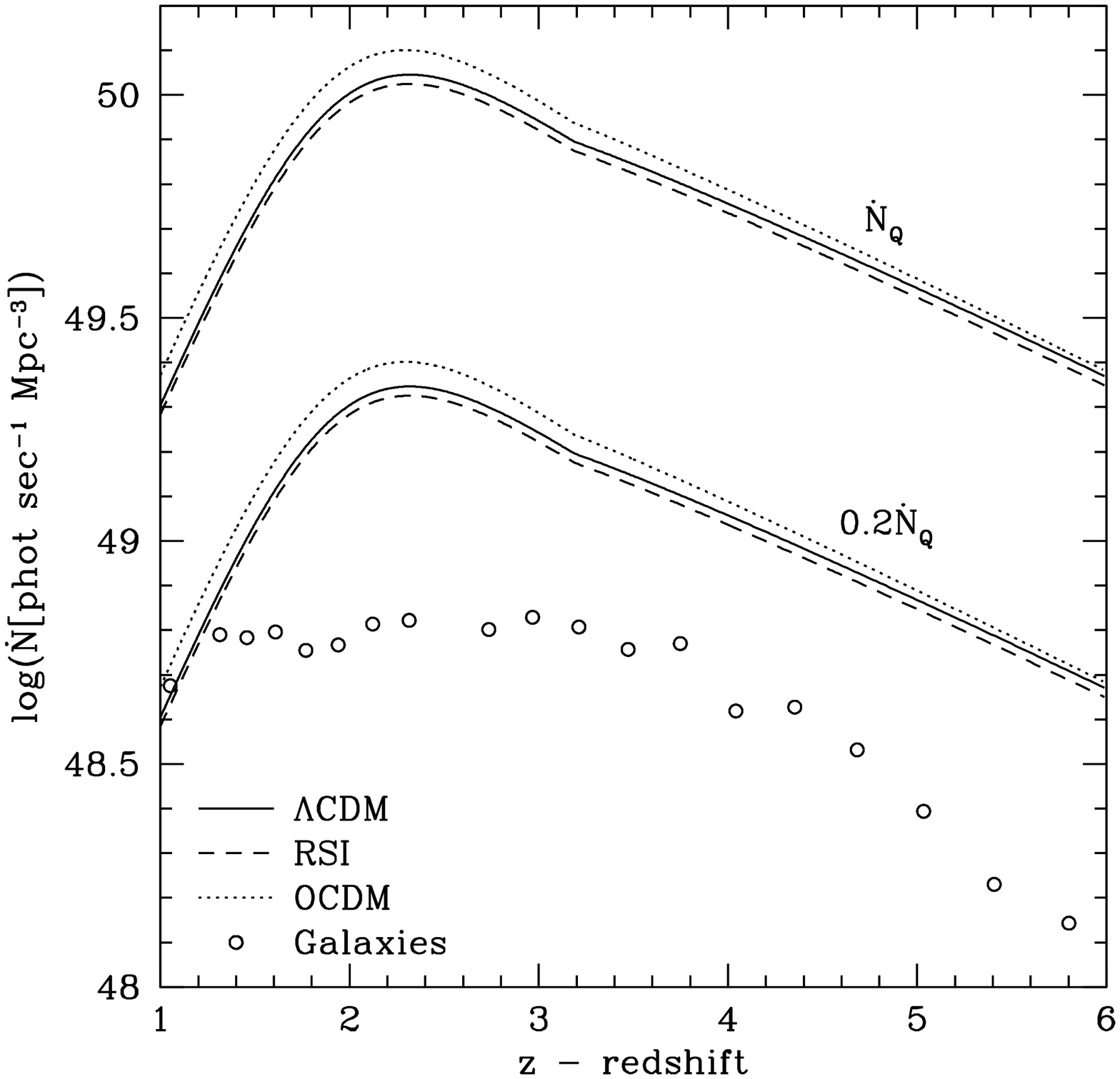,height=3.2in,width=3.2in}}
\caption{The QSO and galaxy emission rate of \heii\ ionizing photons
per unit comoving volume. For QSOs we plot both the full value and
20\% of the emission rate and for three different cosmologies. The
solid line is for $\lcdm$ with $\omg0=0.27$, $\omgl=0.73$ \& $h=0.72$
(Spergel \etal 2003). The dashed line is for the running spectral
index model of Spergel \etal (2003) with $\omg0=0.268$, $\omgl=0.732$
\& $h=0.71$. The dotted line is for OCDM with $\omg0=0.3$, $\omgl=0$
\& $h=0.7$. For galaxies we plot the emission rate with ionizing
photons escape fraction of 10\% and running spectral index
cosmological model.}
\label{fig:rate}
\end{figure}

We assume that double helium reionization is mainly caused by ionizing
photons emitted by QSOs, and that the contribution of galaxies is
negligible. To justify this assumption we have computed the emission
rates of \heii\ ionizing photons from QSOs and galaxies and plotted
them in figure~\ref{fig:rate}. The calculation of the emission rate of
QSOs is based on that of Madau, Haardt \& Rees (1999; hereafter MHR)
who performed a similar calculation for hydrogen (see
Appendix~\ref{apx:ionphotons}). The emission rate of galaxies is
estimated using the {\sc galform} semi-analytic model of galaxy
formation. Specifically, we adopt the model and parameters of Benson
\etal (2002), except for changes in the cosmological parameters
required to match the cosmological model considered here. The reader
is referred to that paper, and references therein, for a full
description of the model. We choose to ignore the effects of {\hi}
reionization on later star formation\footnote{As shown by Benson \etal
(2002), {\nhi} reionization leads to a suppression of star formation
at later times.} so as to obtain an upper limit on the number of
ionizing photons produced at a given redshift. Note that this model
tends to produce more photons than the more recent implementation of
the Durham model presented in Baugh \etal (2004; see their Figure
1). As such, we are overestimating the number of ionizing photons
produced. We assume that 10\% of the produced ionizing photons escape
the galaxies into the IGM (observations of low and high redshift
galaxies imply escape fractions of $\lsim$10\%; Leitherer et al. 1995;
Tumlinson et al. 1999; Steidel, Pettini \& Adelberger 2001). According
to figure~\ref{fig:rate} the emission from QSOs is dominant at
redshifts $z\!\!\gsim\! 1.4$ even if only 20\% of the QSOs emission
calculated value reaches the IGM.

At some high redshift QSOs begin ionizing \heii. The fraction of
\heiii\ prior to that redshift is assumed to be zero.  The mean free
path, $\lph$, of an \heii\ ionizing photon is $\lph\sim 0.8
n_{\nheii,-6}^{-1}{\rm Mpc}$ where $n_{\nheii,-6}$ is the physical
\heii\ number density in $10^{-6}\rm cm^{-3}$. Initially the mean free
path, $\lph$, is much smaller than the mean distance, $d_{\rm QSO}$
between QSOs and so only isolated patches (bubbles) of space are
ionized. As the QSO activity increases more photons become available
and the volume filling factor of ionized bubbles approaches unity. At
this stage, the mean free path in the moderate density IGM, away from
discrete absorbers, can exceed the Hubble radius, the radiation field
becomes nearly uniform, and an equilibrium between recombinations and
photo-ionization is established. This situation is maintained as long
as enough photons are produced.

When a given point in the IGM is ionized, its temperature increases by
an amount proportional to the fraction of \heii. This is a major
source of heating in the IGM as each helium atom can be ionized and
subsequently recombine several times (Miralda-Escud\'{e} \& Rees
1994). The dominant cooling mechanism is adiabatic cooling as a result
of the cosmic expansion (e.g. Miralda-Escud\'{e} \& Rees 1994). During
the patchy period of ionization large temperature differences between
ionized and neutral regions can develop, spoiling any tight relation
between the temperature and the density (T-$\rho$) in the IGM
(e.g. Theuns \etal 1998). If this period is short lived adiabatic
cooling and photo-heating in the later evolution of the IGM work to
re-establish the tight T-$\rho$ relation. Much of the thermal history
of the IGM at $2<z<5$ is governed by the way \heii\ reionization
proceeds.

\section{The Monte-Carlo model} 
\label{sec:model}

\begin{figure}
\centering
\mbox{\psfig{figure=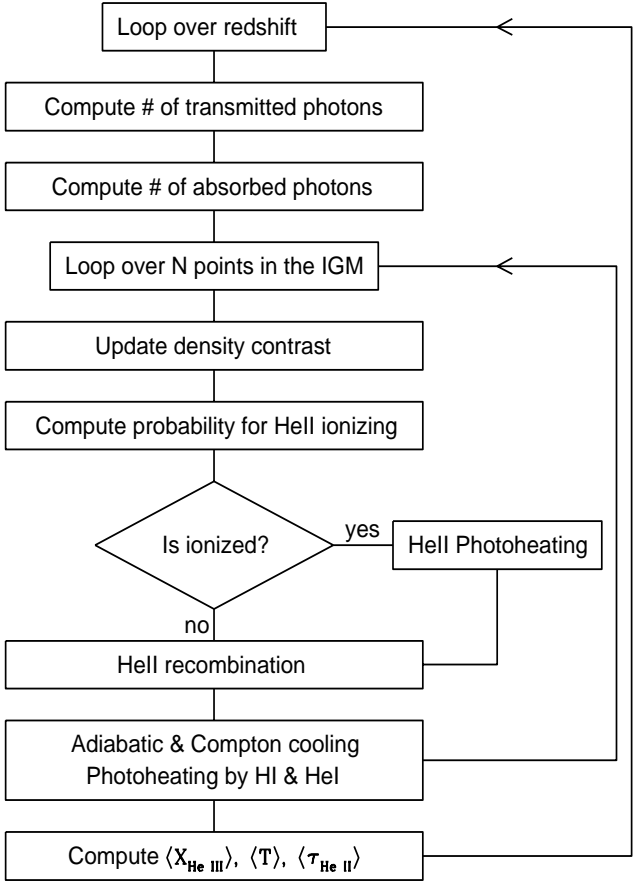,height=5.0in,width=3.6in}}
\caption{Flowchart of the Monte-Carlo method.}
\label{fig:block}
\end{figure}

Given $N$ random points in the IGM with specified gas density,
temperature, and ionization state at some high redshift, $z_{i}$, the
model is designed to yield the corresponding quantities at any lower
redshift. The model assumes that a uniform ionizing background for
hydrogen and singly ionized helium has already been established prior
to $z_{i}$. The model aims at following the physical state of the IGM
as a result of \heii\ reionization by the QSOs population. At the
early stages of \heii\ reionization the mean free path for the
absorption of an \heii\ ionizing photon, $\lph$, is short and the
reionization proceeds in a patchy way as a result of the discrete
distribution of QSOs. At these stages a QSO generates an almost fully
ionized bubble as the bounding thin (thickness $\lph$) ionization
front propagates outward. The process continues until the QSO fades
away. The mean lifetime of a QSO is $\sim 10^7$ years (e.g. Hosokawa
2002; Schirber \etal 2004; Porciani \etal 2004) which is much shorter
than the recombination timescale (order of 1 Gyr at $z=3$) inside the
bubbles.  Therefore, we assume that the QSO radiation is emitted in
bursts. At later stages when the filling factor of \heiii\ regions
becomes significant, the mean free path, $\lph$, becomes so large that
a non-negligible fraction of the absorbed photons at a given time have
actually been produced at significantly larger redshifts. We start the
ionization of \heii\ by QSOs at redshift $z_{i}\approx 6$, with
hydrogen and helium already singly ionized. We neglect the \heii\
ionization by galaxies. Figure \ref{fig:block} presents a schematic
description of the Monte-Carlo model.  The details of the model are as
follows:

\begin{enumerate}

{\bf \item Initialization}: we work with $N$ points representing
random uncorrelated points in the IGM (we use $N=2\times 10^{5}$). At
each point we determine the \heiii\ abundance fraction, $X$, the
temperature, $T$, and the density perturbation, $\delta$. At the
initial time, we set $X=0$ for all the points and assign temperatures
according to $T=T_0(1+\delta)^{2/3}$. At each point, the initial
density contrast $\delta_i^{\rm in}$ at some very high redshift,
$z^{\rm in}=20$, is drawn randomly from a normal (Gaussian)
probability distribution function with zero mean and with rms value
$\sigma_0$.

{\bf \item The local density and its evolution}: the ionized fraction
and the adiabatic cooling/heating depend on the local gas density and
its evolution in time. Given the initial mass power spectrum of the
dark matter we would like to have a recipe for deriving the gas
density contrast as a function of time. Hydrodynamical simulations
strongly support the picture in which the gas distribution in the IGM
traces the dark matter density smoothed over the comoving Jeans length
scale, $\xj$, at which pressure forces roughly balance gravity
(Hernquist \etal 1996; Theuns \etal 1998). We therefore estimate the
gas density power spectrum by smoothing that of the dark matter on the
Jeans length scale (see Appendix~\ref{apx:gasdensity} for details).
We approximate the density contrast $\delta_{i}(z)$ at a point $i$ as
follows. The first step is to obtain the gas density variance
$\left<\delta^{2}_{i}(z)\right>=\sigma^{2}(z)$ as a function of
redshift.  Given the linear power spectrum we use the recipe described
in Peacock (1999) to derive the corresponding nonlinear
power-spectrum. This power-spectrum is then multiplied by an
appropriate Jeans smoothing window and integrated to yield $\sigma^2$.
Let $z^{\rm in}$ be a sufficiently high redshift such that
$\sigma^{2}(z^{\rm in})\ll 1$ and such that the QSO emission is
negligible at $z^{\rm in}$. Draw a random variable $\delta_i^{\rm G}$
from a Gaussian distribution of zero mean and variance
$\sigma^{2}(z^{\rm in})$.  We write the density contrast,
$\delta_{i}(z)$, at any redshift $z<z^{\rm in}$ as
\begin{equation}
\left(1+\delta_i(z)\right)={\rm e}^{\tilde D(z)\delta_i^{\rm G}+\tilde D^2(z)b_0},
\label{eq:lndensity}
\end{equation}
where the nonlinear growth factor, $\tilde D(z)$, is
\begin{equation}
\tilde D(z)=\sqrt{\frac{\ln(1+\sigma^2(z))}{\ln(1+\sigma^2(z^{\rm in}))}}.
\label{eq:lndz}
\end{equation}
The form (\ref{eq:lndensity}) is based on the log-normal model
(e.g. Kofman \etal 1994) for the evolution of the density contrast,
except that in the standard log-normal model the factor $\tilde D$ is
replaced with the linear growth factor of density perturbations. The
form (\ref{eq:lndz}) for $\tilde D$ ensures that the variance of
$\delta_{i}(z)$ is $\sigma^{2}(z)$ which can be estimated reliably
using the recipe outlined below.  On can show that
$\left<\delta^{2}_{i}(z)\right>=\sigma^{2}(z)$, as follows. If
$\delta_{i}^{\rm G}$ is Gaussian, it is easy to see that the mean of
$\delta_{i}(z)$ is zero and the variance is $\exp\left[\tilde
D(z)^{2}\left<(\delta_{i}^{\rm G})^{2}\right>\right]-1$. Hence, using
$\left<(\delta_{i}^{\rm G})^{2}\right>=\sigma^{2}(z^{\rm in})\ll 1$
and (\ref{eq:lndz}), we find that
$\left<\delta^2_i(z)\right>=\sigma^2(z)$. Therefore, the form
(\ref{eq:lndensity}) guarantees a variance of $\sigma^{2}(z)$ for
$\delta_{i}(z)$, provided that
\begin{equation}
b_0=-\frac{1}{2}\sigma_0^2=-\frac{1}{2}\sigma^2(z^{\rm in}).
\label{eq:lnb0}
\end{equation}

{\bf \item The ionization probability}: at each time step the
ionization state of helium is updated as follows. We assign a
probability, $f_{i}$, for photo-ionizing \heii\ at the point $i$, and
draw a random number, $p_i$, between 0 to 1 from a uniform
distribution. If $p_i<f_i$ then the point $i$ is ionized. We adopt the
following form for the photo-ionization probability,
\begin{equation}
f_i=\frac{\nabs}{n_{\nhe}}\frac{N\ P_{\rm bias}(\delta_i)}{\sum_{j=1}^N(1-X_j)(1+\delta_j)P_{\rm bias}(\delta_j)},
\label{eq:probfi}
\end{equation}
where $n_{\nhe}$ is the mean comoving number density of all helium
species, $X_j$ is the \heiii\ abundance fraction in point $j$, and
$\nabs(z)$ is the total number of ionizing photons per unit comoving
volume that are actually absorbed in the current redshift interval
$[z(t),z(t+\Delta t)]$ (see below).  At the beginning of each time
step the number of available ionizing photons is computed as $N_{\rm
trs} + \nq \Delta t $ where $N_{\rm trs}$ is the number photons that
have been transmitted from previous time steps and $ \nq \Delta t $ is
the number of photons that are produced by QSO in the current time
steps. The calculation of $\nq$ is detailed in
Appendix~\ref{apx:ionphotons}. The number of available ionizing
photons is tabulated in frequency bins and $\nabs$ is computed in a
self-consistent way as described in Appendix~\ref{apx:Nabs}.

The factor $P_{\rm bias}$ takes account of the bias between the gas
density distribution and the distribution of the \heii\ ionizing
photons.  At the initial stages of \heii\ reionization, the ionizing
photons are preferentially absorbed near the sources. As \heii\
reionization proceeds, the mean free path, $\lph$, becomes large and
the ionizing radiation eventually forms a uniform background.  In this
case the distribution of the ionizing photons is independent of
$\delta$.  Here we adopt the following form for $P_{\rm bias}$,
\begin{equation}
P_{\rm bias}(\delta_i)\propto (1+\delta_i)^{(b-1)s},
\label{eq:Pbias}
\end{equation} 
where $b$ is the bias parameter between the distribution of the QSOs and
the gas density field, and $s=s(\lph)$ is in general a function of the
mean free path, $\lph$, of an \heii\ ionizing photon. We work with
three choices for $s(\lph)$,
\begin{subequations}
\begin{equation}
s_{\rm const}=1 \; ,
\label{eq:sconst}
\end{equation}
\begin{equation}
s_{\rm linear}=\left\{
\begin{array}{cl}
\xj/\lph, & {\rm if}\ \lph>\xj \\
1, & {\rm otherwise}
\end{array}\right. \; ,
\label{eq:slinear}
\end{equation}
and
\begin{equation}
s_{\rm exp}=\left\{
\begin{array}{cl}
e^{(1-\lph/\xj)}, & {\rm if}\ \lph>\xj \\
1, & {\rm otherwise}
\end{array}\right. \; .
\label{eq:sexp}
\end{equation}
\label{eq:scalefunc}
\end{subequations}

{\bf \item He\ {\footnotesize \bf II} photo-ionization heating}: once
a point is flagged for ionization we update its ionized fraction and
temperature according to,
\begin{equation}
X_i \longrightarrow 1,
\label{eq:fullion}
\end{equation}
and
\begin{equation}
T_i \longrightarrow T_i+\Delta T(1-X^{\rm prev}_i),
\label{eq:heating}
\end{equation}
where $\Delta T$ is the increment in the temperature due to
photo-ionization heating, and $X^{\rm prev}_i$ is the ionization
fraction before the current ionization. We also make an attempt at
modelling the effect of a finite QSO lifetime, $\tq$, on the
reionization process. The model does not contain explicit information
on the spatial distribution of the sources. The way we alleviate this
problem is by assuming that each freshly ionized point remains ionized
for a period of time equal to $\tq$ if it has not been selected for
ionization during this time. During this period of time, for each time
step, some recombination and photo-ionization heating occur, and can
be treated as a local equilibrium.

The process of drawing random numbers for points continues until all
$\nabs$ photons are absorbed.

{\bf \item He\ {\footnotesize \bf II} recombination}:
recombination reduces the \heiii\ abundance fraction by
\begin{equation}
\frac{\dd X (1+\delta)}{\dd t}=-\aheii (1+z)^3\ne X(1+\delta)^{2},
\label{eq:Xevolution}
\end{equation}
where $\aheii$ is the \heii\ case B recombination coefficient
(e.g. Verner \& Ferland 1996) and $\ne$ in the mean comoving number
density of electrons.

At every time step we approximate the reduction in the \heiii\
abundance fraction at every point $i$ due to recombination
\begin{equation}
X_i \longrightarrow X_i\exp\left[-\aheii(T_i)(1+z)^3(1+\delta_i)\ne\Delta t\right].
\label{eq:recrate}
\end{equation}

{\bf \item Adiabatic cooling/heating}: the main cooling process in the
moderate density IGM at redshifts $z\lsim 6$ is adiabatic cooling.
The adiabatic cooling/heating is calculated from the thermodynamic
dependence between the temperature and density of a monatomic ideal
gas
\begin{equation}
\frac{T}{T_0} = \left(\frac{\langle\rho\rangle}{\langle\rho_0\rangle}\right)^{2/3},
\label{eq:tempdens}
\end{equation}
where $T_0$ and $\langle\rho_0\rangle$ are the temperature and the
mean density at some initial redshift, $z_0$. The mean density,
$\langle\rho\rangle\sim (1+z)^{3}$ so that
\begin{equation}
T = T_0\left(\frac{1+z}{1+z_0}\right)^2\left(\frac{1+\delta}{1+\delta_0}\right)^{2/3}.
\label{eq:tempadiabatic}
\end{equation}

Therefore, from equation (\ref{eq:tempadiabatic}), the adiabatic
cooling/heating at every time step is
\begin{eqnarray}
T_i(z) & = & T_i(z+\Delta z)\times\nonumber \\
& & \left(\frac{1+z}{1+z+\Delta z}\right)^2\left(\frac{1+\delta_i(z)}{1+\delta_i(z+\Delta z)}\right)^{2/3}.
\label{eq:cooling}
\end{eqnarray}

{\bf \item Photo-ionization heating by H\ {\footnotesize \bf I} \& He\
{\footnotesize \bf I}}: our basic assumption is that at redshifts
$z\lsim 6$ hydrogen is fully ionized and helium is singly ionized. We
also assume a local thermal equilibrium between the background
radiation and recombination at every time step. For simplicity we
treat \hei\ as extra \hi\ particles, and use the Miralda-Escud\'{e} \&
Rees (1994) analysis for the IGM heating by the \hi\ photo-ionization
background radiation, to calculate the temperature increment at every
point $i$
\begin{equation}
T_i \longrightarrow T_i+\frac{2}{3k}E\ahi(T_i)(1+z)^3(1+\delta_i)\ne\Delta t,
\label{eq:tempHrate}
\end{equation}
where $E$ is the average energy of the absorbed photons minus the
average energy lost per recombination, and $\ahi$ is the \hi\ case B
recombination coefficient (e.g. Verner \& Ferland 1996).

{\bf \item Compton cooling}: Compton cooling resulting from free
electrons scattering off the CMB photons is non-negligible at
redshifts $3<z<6$. This cooling timescale is given by (Peebles 1968)
\begin{equation}
t_{\rm Compton}=\frac{1161.3(1+X_e^{-1})}{(1+z)^4[1-T_0^{\rm CMB}(1+z)/T_e]}{\rm Gyr},
\label{eq:tcompton}
\end{equation}
where $X_e$ is the fraction of free electrons, $T_0^{\rm CMB}$ is the
temperature of the CMB at the present day, and $T_e$ is the
temperature of the free electrons. At every time step the temperature
decrease at every point $i$ due to Compton cooling is
\begin{equation}
T_i \longrightarrow T_i-\frac{T_i\Delta t}{t_i^{\rm Compton}}.
\label{eq:tempcomp}
\end{equation}

{\bf \item Other cooling processes}: we have assessed the importance
of the following cooling processes: collisional ionization,
recombination, dielectronic recombination, collisional excitation, and
bremsstrahlung (e.g. Theuns \etal 1998). We have found that all of
these processes are negligible for $\delta\lsim 100$ in the
temperature range of interest to us $T\lsim 10^{5}$K. One should also
keep in mind that we assume an instantaneous photo-ionization and
photo-heating (equation \ref{eq:fullion}). Rapid energy loss in line
excitations and collisional ionization immediately follows and brings
the temperature to $T\lsim 10^{5}\rm K$. This rapid cooling is
implicitly included in the temperature increment $\Delta T$ in
equation (\ref{eq:heating}). The parameter $\Delta T$ is treated as a
free parameter in our model.

\end{enumerate}

\subsection{The model input}
\label{sec:modelinput}

\begin{table}
\caption{Parameters of the three cosmological models considered here.
The parameters of the power-law $\lcdm$ and the running spectral index
$\lcdm$ models are taken from Spergel \etal (2003). The open CDM model
has zero dark energy component. The parameters $\omg0$, $\omgl$ and
$\Omega_{\rm b}$ are, respectively, the mean density (in units of the
critical value) of the of the dark matter, the dark energy
(cosmological constant) and the baryon component. The Hubble constant,
$h$, in units of $100$ km/s/Mpc, $n$ is the power spectrum power-law,
and $\sigma_8$ is the rms value of density fluctuations in spheres of
radius $8\hmpc$. The spectral index of the RSI model is a function of
wavenumber, $k$, and is given by $n(k)=n(k_0)+[{\rm d}n/{\rm d}\ln
k]\ln(k/k_0)$ with $k_0=0.05$ Mpc$^{-1}$ and ${\rm d}n/{\rm d}\ln
k=-0.031$.}
\vspace{1mm}
\begin{center}
\begin{tabular}{lcccccc}
\hline
& $\omg0$ & $\omgl$ & $\omgb$ & $h$ & $n$ & $\sigma_8$ \\
\hline
\hline
$\lcdm$ & 0.270 & 0.730 & 0.0463 & 0.72 & 0.99 & 0.90 \\
RSI & 0.268 & 0.732 & 0.0444 & 0.71 & Var. & 0.84 \\
OCDM & 0.300 & 0.000 & 0.0463 & 0.70 & 1.00 & 0.90 \\
\hline
\end{tabular}
\end{center}
\label{tbl:cosmogony}
\end{table}

A basic input of the model are the cosmological parameters. We will
present results for three variants of the Cold Dark Matter (CDM)
cosmological model. The first is the $\lcdm$ cosmological model of a
flat universe with a cosmological constant and a power-law power
spectrum of density fluctuations. The second is also a $\lcdm$ model
with similar cosmological parameters but with a running spectral index
(RSI) for the power spectrum. The third is an open CDM (OCDM) model,
similar to our $\lcdm$ model but with no cosmological constant
term. The main parameters defining these cosmologies are listed in
Table~\ref{tbl:cosmogony}. The parameters of the $\lcdm$ and RSI
cosmologies are taken from the first year WMAP data analysis (Spergel
\etal 2003). The OCDM model is not currently viable in view of the
WMAP measurements and the observations of Type Ia SNe (e.g. Knop \etal
2003; Riess \etal 2004). Nevertheless, we run our Monte-Carlo model
with this cosmological model for the sake of comparison.

\begin{table}
\caption{The input parameters of the Monte-Carlo model: $f_{\nq}$ is
the QSO emission rate factor, $\tq$ is the mean QSO lifetime,
$T_0(z\!\!=\!\!6)$ is the initial mean temperature at redshift $z=6$,
$\Delta T$ is the temperature increment induced by photo-ionization of
\heii, $E$ is the cross section weighted excess photon energy for
hydrogen ionization, $\xj(z\!\!=\!\!3)$ is the comoving Jeans length
at redshift $z=3$, $b$ is the QSOs distribution bias parameter, and
$s$ is the scaling function between the QSOs and the ionizing emission
distributions.}
\vspace{1mm}
\begin{center}
\begin{tabular}{lcl}
\hline
Parameter & Values & Units \\
\hline
\hline
$f_{\nq}$ & 0.2, 0.3, 0.4, 0.5, 1.0 & \\
$\tq$ & 0, 5, 10, 20, 40, 60 & Myr \\
$T_0(z\!\!=\!\!6)$ & 1.0, 2.0, 3.0, 4.0, 5.0, 6.0 & $10^4$K \\
$\Delta T$ & 1.0, 1.5, 1.84 & $10^4$K \\
$E$ & 1.27, 2.54, 3.81 & $\ev$ \\
$\xj(z\!\!=\!\!3)$ & 0.05, 0.1, 0.2, 0.3, 0.4 & $\hmpc$ \\
$b$ & 1.0, 2.0, 3.0, 5.0, 8.0 & \\
$s$ & const, linear, exp & \\
\hline
\end{tabular}
\end{center}
\label{tbl:compare}
\end{table}

The Monte-Carlo model has also to be fed with the input parameters
pertaining to the physical processes discussed previously. These
parameters are summarized in Table~\ref{tbl:compare}. The first
parameter in this table, $f_{\nq}$, is the QSO emission factor. In
Appendix~\ref{apx:ionphotons} we describe our calculation of the QSO
emission rate of ionizing photons per unit comoving volume,
$\nq(z)$. The calculation is based on the MHR luminosity
function. Although we work with the basic shape of the inferred
$\nq(z)$ we introduce the factor $f_{\nq}$ to allow for an overall
reduction in the intensity of emitted radiation. The motivation behind
this factor is that several observations and simulations suggest a
softer UV radiation field than the standard UV background due to QSOs
emission (MHR). Smette \etal (2002) has shown regions of the HE
2347-4342 spectrum in which the \heii\ optical depth is very high and
the corresponding \hi\ optical depth is low required photo-ionization
by a very soft UV radiation field. Independent evidence is provided by
the column density ratios of highly ionized metals ($N_{\rm Si\
IV}/N_{\rm C\ IV}$ versus $N_{\rm Si\ II}/N_{\rm C\ IV}$) at high
redshifts (Giroux \& Shull 1997; Savaglio \etal 1997; Songaila 1998;
Kepner \etal 1999). Numerical simulations of the Ly$\alpha$ forest
also suggest a softer UV radiation field (Theuns \etal 1998;
Efstathiou \etal 2000).

Next is $\tq$, the mean QSO lifetime. Since the points in the model
are not spatially correlated we can take into account the QSOs
lifetime only in a very limited way. We assume that a point that was
ionized will stay ionized at least for the QSO lifetime. Therefore,
$\tq=0$ merely implies that the QSO lifetime is shorter then the model
time step.

The mean temperature at redshift $z=6$, $T_0(z\!\!=\!\!6)$, is fixed
prior to \heii\ reionization. This temperature is set by the interplay
between cooling (mainly adiabatic and Compton) and the photo-heating
by the hydrogen ionizing background.

The parameter $\Delta T$ is the increment in the temperature due to
\heii\ reionization (cf. equation \ref{eq:heating}). In the limit of a
very high ionization fraction an absorbed photon produce only one free
electron which loses its energy by Coulomb interactions in the ionized
IGM. The temperature increment in this case is given by
(Miralda-Escud\'{e} \& Rees 1994)
\begin{equation}
\Delta T=\frac{2}{3k}X\left(<\!\!\epsilon\!\!>-\ \epsilon_0\right)\simeq
1.84\times10^4{\rm K},
\end{equation}
where $X=n_{\rm He}/(n_{\rm H}+n_{\rm He}+n_{\rm e})\approx 0.07$ is
the fraction of particles which participate in the photo-ionization
process from the total number of particles in the gas,
$<\!\!\epsilon\!\!>\ \simeq 1.96\times 10^{-10}{\rm erg}$ is the mean
energy per photon of photons with energies greater than $4\ {\rm Ry}$
which are emitted by QSOs, and $\epsilon_0=4\ {\rm Ry}\simeq
8.72\times 10^{-11}\ {\rm erg}$ is the energy spent per \heii\
ionization.

However, when the ionization fraction is not very high, the electrons
produced by photo-ionization may have a large probability of
interacting with \hei\ and lose energy in line excitations and
collisional ionization. Furthermore, when secondary electrons are
produced, the initial energy must be shared among several electrons,
and so the final energy per electron is reduced. Therefore, we also
explore lower values of $\Delta T$.

The cross section weighted photon energy, $E$, determines the
photo-heating due to the \hi\ and \hei\ ionizing photon backgrounds,
assuming local equilibrium between the background radiation and the
recombination processes (equation \ref{eq:tempHrate}). Following
Miralda-Escud\'{e} \& Rees (1994) we neglect the temperature
dependence of $E$ (which is valid for low temperatures, since the
energy lost in recombination is then negligible). In the case of pure
hydrogen and for QSOs SED $L(\nu)\propto\nu^{-1.5}$,
Miralda-Escud\'{e} \& Rees (1994) found $E=0.28\ {\rm
Ry}=3.81\ev$. Since $E$ should also include the \hei\ ionizing photon
background and since we use $L(\nu)\propto\nu^{-1.8}$ (see
Appendix~\ref{apx:lumspec}), we explore lower values of $E$ as well.

The next parameter, $\xj(z\!\!=\!\!3)$, is the Jeans length (comoving)
at redshift $z=3$\footnote{Although the Jeans length is in principle
specified by the temperature of the IGM we choose to treat it here as
a free parameter of the model. The shape of the Jeans filtering window
is quite uncertain---even in the linear regime, it depends
non-trivially on the reionization history (e.g. Hui \& Gnedin 1997,
Nusser 2000). The situation becomes more complicated when nonlinear
effects become important.}. The Jeans length, $\xj$, is defined as the
scale over which the dark matter density should be smoothed to yield
an estimate of the gas density (see
Appendix~\ref{apx:gasdensity}). Large values of $\xj$ yield smaller
fluctuations, which in turn decreases the recombination rate, and also
affects the adiabatic cooling. We assume a simple evolution of $\xj$
which depends only on redshift
\begin{equation}
\xj(z)=\xj(z\!\!=\!\!3)\left(\frac{4}{1+z}\right)^{1/2}.
\label{eq:XJeans}
\end{equation}

The QSOs bias parameter, $b$, determines the QSOs distribution from
the gas density distribution, and the scaling function, $s$, scales
the ionizing emission distribution from the QSOs distribution. We used
three different scaling functions constant, linear and exponential
(see equations \ref{eq:sconst}, \ref{eq:slinear} \& \ref{eq:sexp}).


\section{Results}
\label{sec:compare}

\begin{figure*}
\resizebox{0.83\textwidth}{!}{\includegraphics{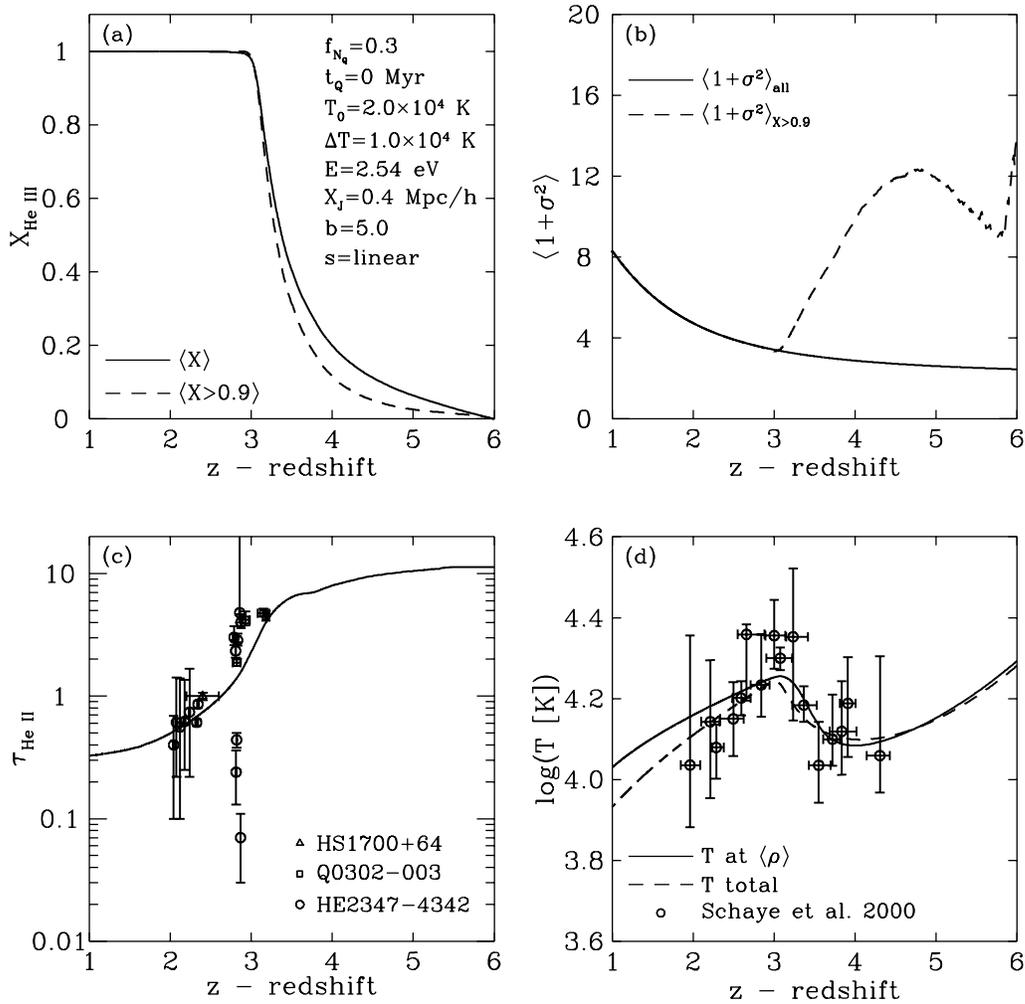}}
\caption{Results for the power-law $\lcdm$ cosmological model for a
choice of set of parameters as indicated in the figure. This set of
parameters gives $\chi_T^2=0.76$ and $\chi_\tau^2=0.75$. {\bf Panel
(a)}: The \heiii\ fraction (see text) as a function of redshift. The
solid line is the mean fraction of \heiii. The dashed line is the
volume filling factor of regions with \heiii\ fraction above 0.9. {\bf
Panel (b)}: The mean clumping factor, $\langle 1+\sigma^2\rangle$, as
a function of redshift. The solid line is the mean clumping
factor. The dashed line is the clumping factor for regions with
\heiii\ fraction above 0.9. {\bf Panel (c)}: The \heii\ mean optical
depth as a function of redshift. The solid line is the mean optical
depth calculated from the model. The symbols are measurements of the
\heii\ optical depth from the spectrum of the \heii\ Gunn-Peterson
effect towards the quasars HS 1700+64 (triangles, Davidsen \etal
1996), Q0302-003 (squares, Dobrzycki \etal 1991; Heap \etal 2000), and
HE 2347-4342 (circles, Smette \etal 2002; Zheng \etal 2004). The noise
in the line is due to the limited number of points used in the
Monte-Carlo model. {\bf Panel (d)}: The mean temperature as a function
of redshift. The solid line is the mean temperature of the mean
density regions, calculated from regions with $|\delta|<0.2$. The
dashed line is the overall mean temperature, calculated from all
regions. The circles with error bars are the observed mean temperature
calculated from nine quasars Ly$\alpha$ spectra (Schaye \etal 2000).}
\label{fig:wcdmres}
\end{figure*}

\begin{figure*}
\resizebox{0.83\textwidth}{!}{\includegraphics{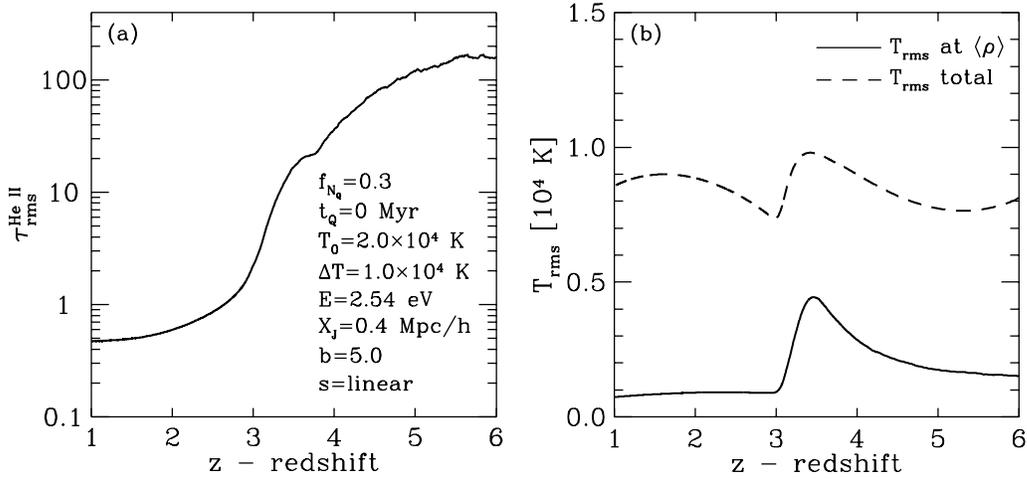}}
\caption{The rms values of optical depth and the temperature
fluctuations in the power-law $\lcdm$ cosmological model for the set
of parameters indicated in the figure. {\bf Panel (a)}: The \heii\
optical depth rms as a function of redshift. The noise in the curve is
due to the limited number of points used in the Monte-Carlo
model. {\bf Panel (b)}: The rms of temperature fluctuations as a
function of redshift. The solid line is the temperature rms in regions
with $|\delta|<0.2$. The dashed line is the overall temperature rms,
calculated from all regions.  The rms value of the fluctuations in a
variable $Z$ is $Z_{\rm rms}=\langle(Z-\langle
Z\rangle)^{2}\rangle^{1/2}$.}
\label{fig:wcdmrms}
\end{figure*}

\begin{figure*}
\resizebox{0.83\textwidth}{!}{\includegraphics{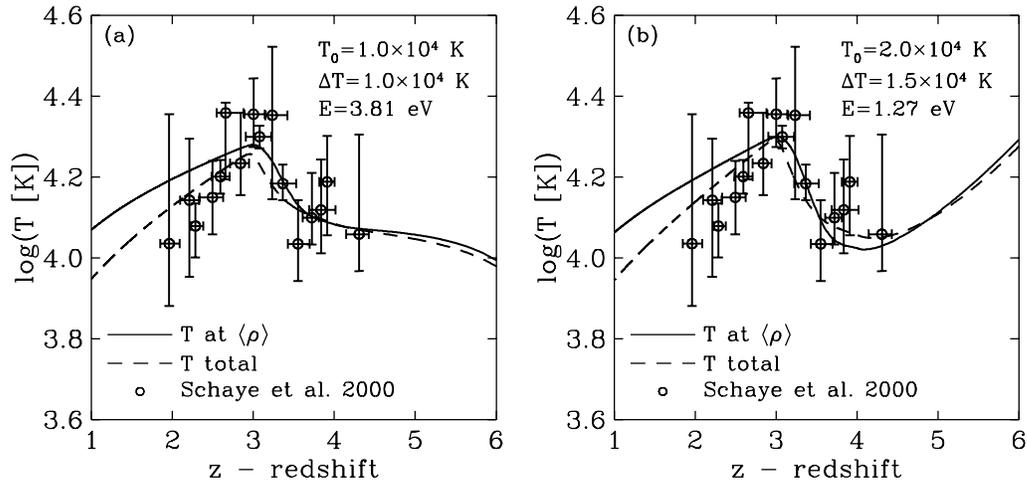}}
\caption{The mean temperature as a function of redshift in the
power-law $\lcdm$ cosmological model for two sets of parameters as
indicated in the panels. The rest of the model parameters are the same
for both cases: $f_{\nq}=0.3$, $\tq=0$ Myr, $\xj(z\!\!=\!\!3)=0.3$,
$b=5$ \& $s=$ linear. The set of parameters of panel (a) gives
$\chi_T^2=0.93$ and $\chi_\tau^2=0.77$. The set of parameters of panel
(b) gives $\chi_T^2=1.0$ and $\chi_\tau^2=0.94$. The solid line is the
mean temperature of the mean density regions, calculated from regions
with $|\delta|<0.2$. The dashed line is the overall mean temperature,
calculated from all regions. The circles with error bars are the
observed mean temperature calculated from nine quasars Ly$\alpha$
spectra (Schaye \etal 2000).}
\label{fig:wcdmtmp}
\end{figure*}

\begin{figure*}
\resizebox{0.83\textwidth}{!}{\includegraphics{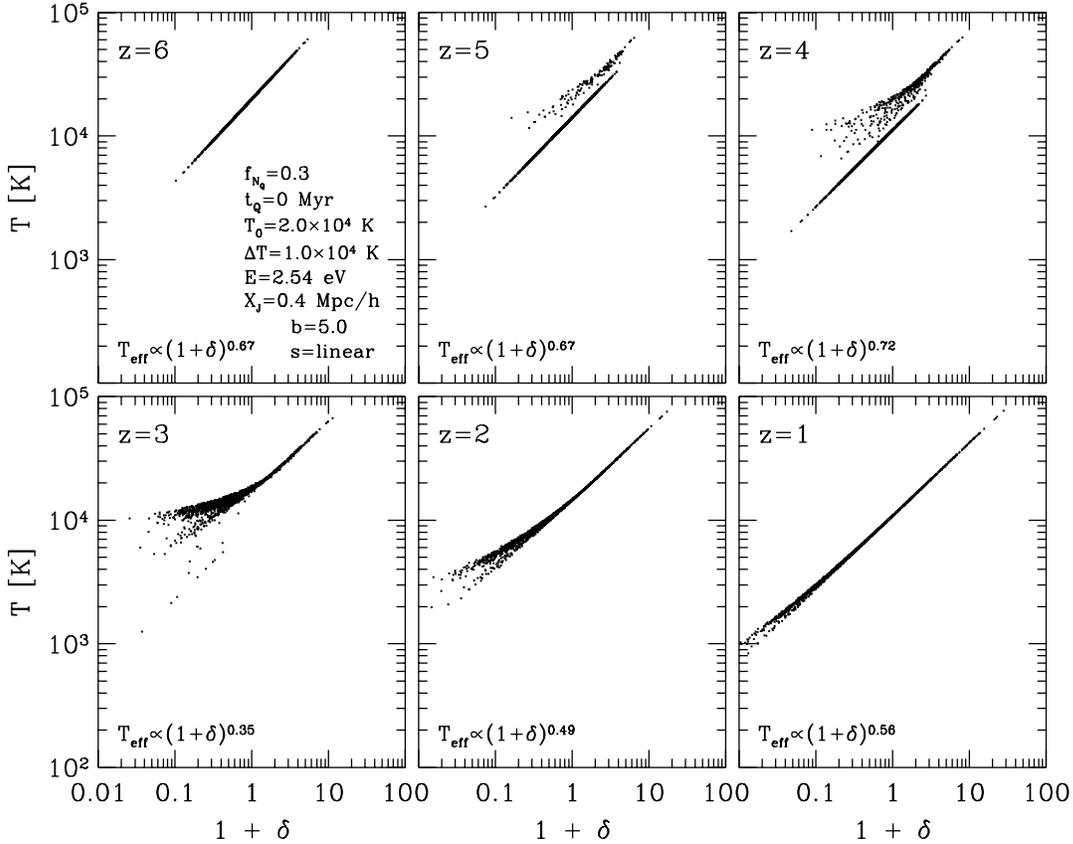}}
\caption{The temperature-density relation for the power-law $\lcdm$
cosmological model. The values of the key parameters are indicated in
the figure. The six panels show the temperature as a function of the
density contrast $\delta$ at different redshifts. The dots are the
local temperature values. $T_{\rm eff}$ is the effective equation of
state assuming $T_{\rm eff}\propto (1+\delta)^{\gamma-1}$ for all
points ionized and non-ionized. The effective temperature is
calculated from the least squares linear fit between $\log(T)$ and
$\log(1+\delta)$.}
\label{fig:wcdmzdt}
\end{figure*}

\begin{figure*}
\resizebox{0.83\textwidth}{!}{\includegraphics{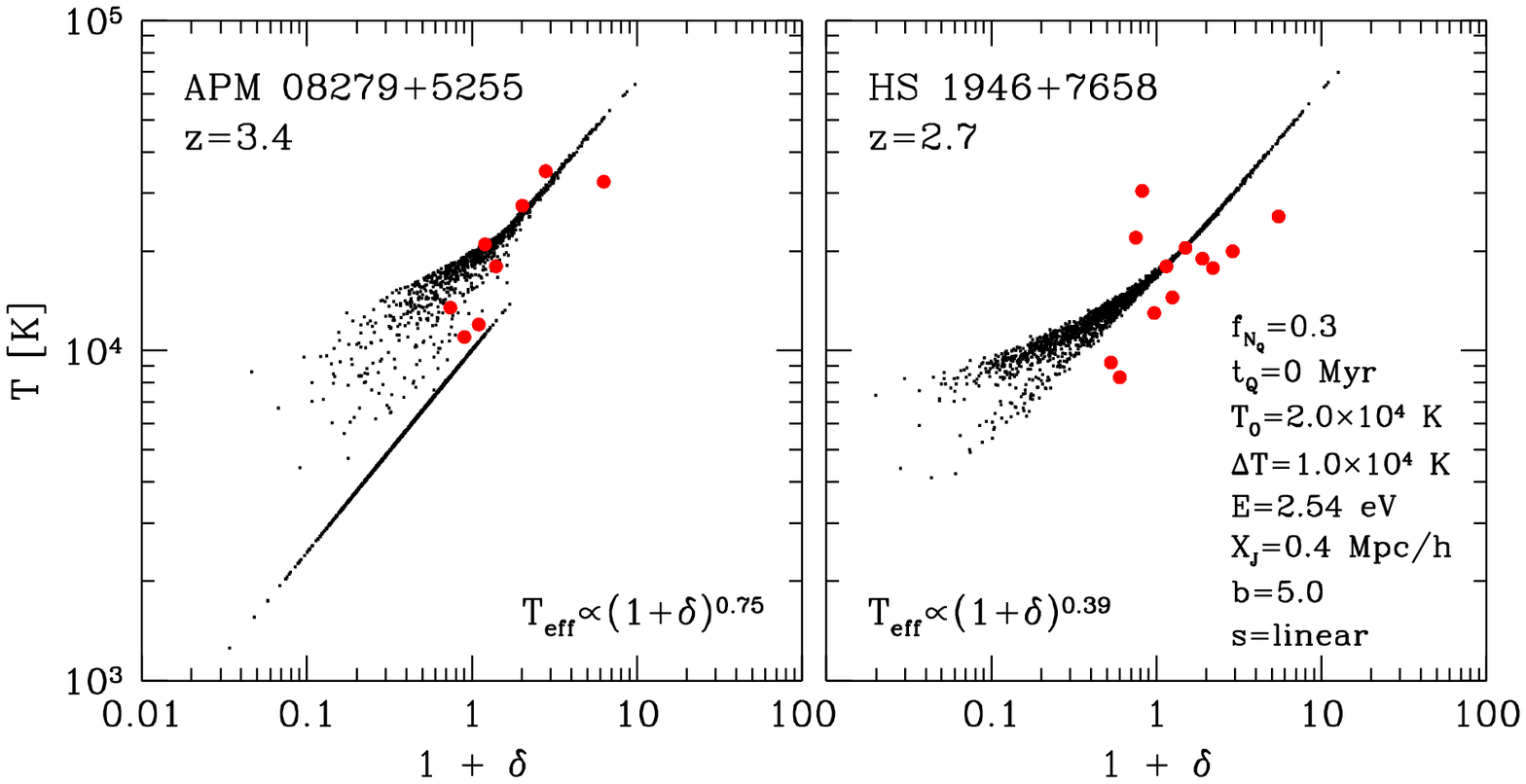}}
\caption{The temperature-density relation as derived from two sets of
observed absorption-line systems (Bryan \& Machacek 2000) and from the
Monte-Carlo model with the same parameters set as in figure
\ref{fig:wcdmzdt}. The effective equation of state, $T_{\rm
eff}\propto (1+\delta)^{\gamma-1}$, is calculated from the least
squares linear fit between $\log(T)$ and $\log(1+\delta)$. {\bf Right
panel}: HS 1946+7658 at $\langle z\rangle=2.7$ (red circles), data
from the Monte-Carlo model at the same redshift (black dots). {\bf
Left panel}: APM 08279+5255 at $\langle z\rangle=3.4$ (red circles),
data from the Monte-Carlo model at the same redshift (black dots).}
\label{fig:wcdmzdt2}
\end{figure*}

We run our Monte-Carlo model for a wide range of the input parameters
listed in Table ~\ref{tbl:compare}.
\renewcommand{\labelenumi}{(\alph{enumi})} For every set of
parameters, the Monte-Carlo model provides:
\begin{enumerate}
\item The \heiii\ actual volume filling factor and the volume filling
factor of regions with \heiii\ fraction above 0.9 as a function of
redshift.
\item The mean clumping factor for all regions and for regions with
\heiii\ fraction above 0.9 as a function of redshift.
\item The mean temperature as a function of redshift for the mean
density regions (regions with $|\delta|<0.2$) and for all regions.
\item The \heii\ mean optical depth as a function of redshift.
\item The IGM equation of state, i.e., the $\trho$ relation.
\end{enumerate}

We compare our results with available observational data. These data
include $(i)$ the mean temperature in regions of densities around the
mean ($|\delta|<0.2$) estimated from nine quasars Ly$\alpha$ spectra
(Schaye \etal 2000), and $(ii)$ the \heii\ mean optical depth,
$\tau_{\nheii}(z)$ measured from the spectrum of the quasars HS
1700+64 (Davidsen \etal 1996), Q0302-003 (Dobrzycki \etal 1991; Heap
\etal 2000), and HE 2347-4342 (Smette \etal 2002; Zheng \etal 2004).

Unfortunately the available observational data are very limited. The
uncertainties in the mean temperature data are large, while the
uncertainties in the mean optical depth data are fairly small. There
is a huge scatter in the optical depth measurements around redshift
$z=2.8$ (i.e. a scatter larger than expected on the basis of the
reported error bars) and it is therefore not meaningful to use these
measurements to compute a mean optical depth. Nevertheless the data
still provide useful constraints on the model input parameters.

As a measure of the ``goodness of fit'' of a given choice of
parameters we use a $\chi^{2}$ statistic. For the temperature data we
define $\chi^{2}_{\rm T}$ as
\begin{equation}
\chi^{2}_{T}=\sum\frac{(T_{\rm data}-T_{\rm model})^{2}}{\sigma_{T}^{2}} \; ,
\end{equation}
where the $T_{\rm data}$ are the measured temperature in the data and
$T_{\rm model}$ are the model predicted temperatures at the same
redshifts as the data. The estimate of the errors in the data,
$\sigma_{T}$, (Schaye \etal 2000) should serve as a general guide
only. We also define a $\chi^{2}$ statistic for the optical depth
measurement. We interpret these data as measurements of the local
optical depth rather than the mean optical depth. Therefore, we write
$\chi^{2}_{\tau}$ as
\begin{equation}
\chi^{2}_{\tau}=\sum\frac{(\tau_{\rm data}-\tau_{\rm model})^{2}}{(\sigma_{\tau_{\rm data}}^{2}+\sigma_{\tau_{\rm model}}^{2})} \; ,
\end{equation}
where the $\tau_{\rm data}$ are the measured optical depth in the data
and $\tau_{\rm model}$ are the model predicted mean optical depths at
the same redshifts as the data. The observational errors in the data
are $\sigma_{\tau_{\rm data}}$ and the scatter about the mean optical
depth as estimated in the model is $\sigma_{\tau_{\rm
model}}=\langle(\tau-\langle\tau\rangle)^2\rangle^{1/2}$ (see figure
\ref{fig:wcdmrms}).

\subsection{Results with a power-law $\Lambda$CDM cosmology}

We first present results for the $\lcdm$ cosmological model (see Table
~\ref{tbl:cosmogony}). We discuss other cosmologies in section
\S~\ref{sec:cosmmodels}.

In general, the best fits of the model to the observations, with
$\chi_T^2\le 1$ and $\chi_\tau^2\le 1$, are obtained for $\fnq\approx
0.3$, Jeans scale of $\xj(z\!\!=\!\!3)\approx 0.2-0.4\hmpc$, and QSOs
lifetime $\tq\lsim 10$ Myr. We found no constrains on the bias
parameter $b$, and the scaling function $s$.

The initial mean temperature is $T_0(z\!\!=\!\!6)\lsim 40000$K. The
temperature increment is $\Delta T\approx 10000-15000$K, which, as
expected, is smaller than the estimate of Miralda-Escud\'{e} \& Rees
(1994) who find $\Delta T\simeq 18400$K for a fully ionized
regions. This is due to energy losses by line excitation and
collisional ionization in partially ionized regions. We found a
correlation between the temperature increment, $\Delta T$, and the
hydrogen photo-ionization heating parameter, $E$. For lower
temperature increment, $\Delta T\approx 10000$K, the heating parameter
is $1.27\lsim E\lsim 3.81\ev$, but for higher temperature increment,
$\Delta T\approx 15000$K, the heating parameter is only $E\approx
1.27\ev$.

Figure \ref{fig:wcdmres} summarizes some of our main results for the
$\lcdm$ cosmology with parameters (listed in the top-left panel) for
which the Monte-Carlo model yields one of the best fits to the
observational data, $\chi_T^2=0.76$ and $\chi_\tau^2=0.75$.

We quantify the development of \heiii\ regions by computing the mean
ionized fraction, $X$ (shown as the solid curve in panel (a) of figure
\ref{fig:wcdmres}), and the fraction of volume occupied by regions
having $X>0.9$ (dashed curve). The dashed and solid curves behave as
expected. The redshift at which the dashed curve reaches unity in
panel (a) is $z=2.9$.

Panel (b) of figure \ref{fig:wcdmres} shows the evolution of the
clumping factor, $\fclmp\equiv\langle 1+\sigma^2\rangle$, for all
regions (solid curve) and for regions with $X>0.9$ (dashed curve) as a
function of redshift. The clumping factor for all regions increases in
time as expected. The clumping factor indicated by the dashed curve is
large at high redshifts since the reionization starts in dense
regions, and decreases in time until it meets the solid curve around
the redshift where full reionization is established. At later times
all the regions are almost completely ionized, and therefore the
dashed curve follows the solid curve.

Next we examine the optical depth, $\tau_{\nheii}$, for helium
Ly$\alpha$ absorption. This is given by \begin{equation}
\tau_i^{\nheii}=\frac{c\
\sigma_l}{H(z)}n_{\nhe}(1+z)^3(1-X_i)(1+\delta_i),
\label{eq:localoptdepth}
\end{equation}
where $H(z)$ is the Hubble constant at redshift $z$, and
$\sigma_l={\pi e^2}/m_e c^2]\lambda_l f_{ij}\simeq 1.1\times10^{-18}\
{\rm cm^2}$ is the integrated Ly$\alpha$ scattering cross section of
the \heii\ transition at wavelength $\lambda_l=304\Ang$ and oscillator
strength $f_{ij}$. At every time step, we calculate the optical depth
at each point $i$, and the mean optical depth
\begin{equation}
\tau_{\nheii}=-\ln\left[\frac{1}{N}\sum_{i=1}^N\exp\left(-\tau_i^{\nheii}\right)\right]\; .
\label{eq:meanoptdepth}
\end{equation}
The solid line in panel (c) of figure \ref{fig:wcdmres} is
$\tau_{\nheii}$ computed from our model as a function of redshift
according to (\ref{eq:meanoptdepth}). The noise in the line is due to
the limited number of points used in the Monte-Carlo model. For
comparison we also show the available observational data points
measured from the spectrum of the \heii\ Gunn-Peterson effect towards
the quasars HS 1700+64 (triangles, Davidsen \etal 1996), Q0302-003
(squares, Dobrzycki \etal 1991; Heap \etal 2000), and HE 2347-4342
(circles, Smette \etal 2002; Zheng \etal 2004). One can easily see a
significant decrement in the optical depth from $\tau\sim 10$ at
$z\gsim 3.5$ to $\tau\lsim 0.5$ at $z\lsim 2.5$.

In panel (a) of figure \ref{fig:wcdmrms} we see that the rms of the
fluctuations of the optical depth, $\tau_{\rm
rms}=\langle(\tau-\langle \tau\rangle)^{2}\rangle^{1/2}$, behaves in a
similar manner to the mean optical depth. At high redshifts, $z\gsim
3.5$, the high density regions get ionized while the low density
regions stay neutral. This results in a large scatter of the optical
depth ($\tau_{\rm rms}\gsim 20$). The scatter decreases towards full
reionization\footnote{By {\it full reionization} we mean a state in
which the volume filling factor of regions with $X>0.9$ is unity.}. By
$z\sim 3$, most of the high density regions are already ionized as
well as a large fraction of the low density regions. At $z\lsim 2.5$
we have $\tau_{\rm rms}\lsim 1$.

Panel (d) of figure \ref{fig:wcdmres} shows the evolution of the mean
IGM temperature as a function of redshift. The model predictions are
shown as the solid and dashed curves. In each panel, the solid curve
refers to the mean temperature at points with density contrast
$|\delta|<0.2$, while the dashed line is the volume weighted mean
temperature defined as the average temperature of all the $N$ points
of the Monte-Carlo model. Overlaid as the points with error bars are
the temperatures as inferred from the observations (Schaye \etal
2000). The data points correspond to temperatures in regions of mean
density, and therefore they should be directly compared with the solid
curves. However, the difference between the solid and dashed curves in
each panel is within the error bars. The gradual variation of the
model temperatures with redshift is in reasonable agreement with the
data. Comparing with panel (a), we see that the temperature curves
peak at the redshift for which full reionization is reached. At higher
redshifts each \heiii\ particle can recombine more than once,
resulting in an efficient photo-heating by the \heii\ ionizing
photons. At lower redshifts, the recombination is greatly reduced and,
subsequently, the photo-heating rate decreases and adiabatic cooling
dominates. At these low redshifts, the decline of the solid lines does
not seem to be fast enough to match the observational data. The dashed
curves, however, fall nicely among the data points. This difference
between the dashed and solid curves in each panel is due to the slower
adiabatic cooling of regions with $|\delta|<0.2$, relative to the
overall volume weighted cooling which is dominated by the expanding
low density regions.

Unsurprisingly we found a tight relation between the shape of the mean
temperature curve and the combination of the heating sources $\Delta
T$ and $E$. In the case we have investigated so far, with low
temperature increment $\Delta T=10000$K and moderate hydrogen
photo-ionization heating parameter $E=2.54\ev$, the temperature rises
in $5000-6000$K during $3\lsim z\lsim 4$, and never rises above
18200K. We also present two other typical cases in figure
\ref{fig:wcdmtmp}. In panel (a) of figure \ref{fig:wcdmtmp} we present
a case of low $\Delta T=10000$K and high $E=3.81\ev$. In this case the
rise in the temperature during reionization is around 7500K. The high
hydrogen photo-ionization heating parameter does not allow the IGM to
cool at higher redshifts, $4\gsim z\gsim 6$, and the temperature
reaches a maximum around 19000K. In panel (b) of figure
\ref{fig:wcdmtmp} we present a case of high $\Delta T=15000$K and low
$E=1.27\ev$. In this case the rise in the temperature during
reionization exceeds 8000K and can reach 11000K. The temperature can
reach a maximum of 20000K.

It is instructive to examine the temperature fluctuations in
space. This can be quantified in our Monte-Carlo model by computing
the rms of the temperature fluctuations, $T_{\rm
rms}=\langle(T-\langle T\rangle)^{2}\rangle^{1/2}$. In panel (b) of
figure \ref{fig:wcdmrms} we show the temperature rms as a function of
redshift. The temperature rms of regions with mean density (the solid
lines) is affected mainly by helium photo-heating. When the IGM
reaches full ionization, the photo-heating is not effective any more
and the temperature rms decreases to approximately the same value as
before ionization.

Finally, we consider the effect of $\heii$ ionization on the $\trho$
relation. In the absence of $\heii$ reionization, we expect a tight
$\trho$ relation resulting from adiabatic cooling and photo-heating
(e.g. Hui \& Gnedin 1997). We assume in the Monte-Carlo model that at
$z=6$ the $\trho$ relation is such that $T\propto (1+\delta)^{2/3}$.
To emphasize the effect of patchy helium reionization we do not
include any scatter in the $\trho$ relation at $z=6$.

Figure \ref{fig:wcdmzdt} presents the results for the same set of
parameters as the above. The six panels in the figure, correspond to
the $\trho$ relation at different redshifts. At the bottom of each
panel, an effective equation of state of the form $T_{\rm eff}\propto
(1+\delta)^{\gamma-1}$ is given. This effective equation of state is
derived by the linear least square fit between $\log(T)$ and
$\log(1+\delta)$.

In figure \ref{fig:wcdmzdt} we see that during the early stages of
reionization, when mostly high density regions are being ionized, the
slope of temperature-density relation is steepened and the scatter
increases significantly. At $z\sim 3-4$ a non-negligible scatter of
about $50\%$ is introduced. Around $z=3$, where the reionization of
low density regions is much more significant, the slope of
temperature-density relation flattens. When the helium in the IGM is
fully ionized the photo-ionization heating becomes inefficient and the
main process affecting the gas temperature is adiabatic cooling. The
cooling reduces the scatter, and by $z=1$ a tight $\trho$ relation is
restored.

In figure \ref{fig:wcdmzdt2} we present the $\trho$ relation as
derived from two sets of observed quasar absorption-line systems from
Bryan \& Machacek (2000): APM 08279+5255 at $\langle z\rangle=3.4$
(left panel) and HS 1946+7658 at $\langle z\rangle=2.7$ (right panel),
and compare with the $\trho$ relation from the Monte-Carlo model at
the same redshifts.

At $z=2.7$ the slope of the effective equation of state from the
model, $\gamma_{\rm model}=1.392\pm 0.002$, fits the slope of the data
from the HS 1946+7658 absorption-lines, $\gamma_{\rm data}=1.31\pm
0.15$, within $0.6\sigma$. At $z=3.4$ the slope of the effective
equation of state from the model, $\gamma_{\rm model}=1.748\pm 0.008$,
fits the slope of the data from the APM 08279+5255 absorption-lines,
$\gamma_{\rm data}=1.55\pm 0.14$, within $1.4\sigma$. The coefficient
of the effective equation of state from the model fits the data from
the observations within $1.2\sigma$ at $z=2.7$ and within $0.3\sigma$
at $z=3.4$.

\begin{figure}
\mbox{\epsfig{figure=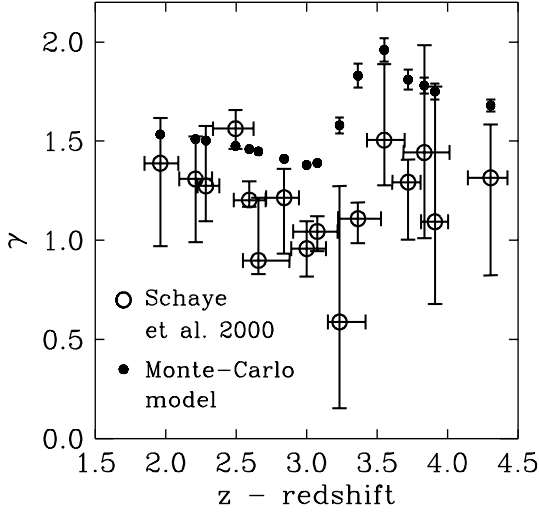,height=3.2in,width=3.2in}}
\caption{The slope, $\gamma$, of the effective equation of state,
$T_{\rm eff}\propto (1+\delta)^{\gamma-1}$, at the mean density. The
open circles are the slopes calculated from the nine QSO Ly$\alpha$
spectra in panel (d) of figure~\ref{fig:wcdmres} by Schaye \etal
2000. The filled circles are the slopes calculated from the
Monte-Carlo model for regions with $|\delta|<0.2$ at the same
redshifts. $1\sigma$ error bars in $\gamma$ are presented for both the
observational data and the model points.}
\label{fig:gamma}
\end{figure}

In figure~\ref{fig:gamma} we present the slopes, $\gamma$, of the
effective equation of state at the mean density of the nine quasar
Ly$\alpha$ spectra studied by from Schaye \etal (2000; open circles)
and the slopes calculated from the Monte-Carlo model from regions with
$|\delta|<0.2$ at the same redshifts (filled circles). As expected,
the reheating of the gas at the time of reionization, $z\sim 3-4$,
causes a decrease in $\gamma$. After the reionization the IGM
gradually evolves again towards an ionization equilibrium and $\gamma$
increases. To compare between the slopes from the observational data
and the slopes from the model we calculated $\eta$, the distance in
$\sigma$ between the two $\gamma$,
\begin{equation}
\eta=\frac{|\gamma_{\rm data}-\gamma_{\rm model}|}{\sqrt{\sigma^2_{\rm
data}+\sigma^2_{\rm model}}} \; ,
\end{equation}
where $\gamma_{\rm data}$ and $\sigma_{\rm data}$ are the
observational data slopes and their $1\sigma$ errors, and $\gamma_{\rm
model}$ and $\sigma_{\rm model}$ are the Monte-Carlo model slopes and
their $1\sigma$ errors (see Table~\ref{tbl:gamma}). From
figure~\ref{fig:gamma} one can see that in general $\gamma$ from the
Monte-Carlo model have higher values then $\gamma$ from the
observational data, although for most redshifts $\eta<3$ (see
Table~\ref{tbl:gamma}).

\begin{table}
\caption{Comparison between the slopes, $\gamma$, of the effective
equation of state at the mean density from the observations (Schaye
\etal 2000) and the slopes from the Monte-Carlo model (from regions
with $|\delta|<0.2$). $z$ is the redshift with the errors from the
observational data. $\gamma_{\rm data}$ is the slope from the
observational data. $\gamma_{\rm model}$ is the slope from the
Monte-Carlo model. $\eta$ is the distance in $\sigma$ between the
model slope and the observational data slope.}
\vspace{1mm}
\begin{center}
\begin{tabular}{cccc}
\hline
$z$ & $\gamma_{\rm data}$ & $\gamma_{\rm model}$ & $\eta$ \\
\hline
\hline
$4.307_{-0.161}^{+0.121}$ & $1.315_{-0.491}^{+0.268}$ & $1.68\pm 0.03$ & $1.35$ \\
$3.911_{-0.098}^{+0.095}$ & $1.094_{-0.415}^{+0.680}$ & $1.75\pm 0.04$ & $0.96$ \\
$3.836_{-0.148}^{+0.179}$ & $1.443_{-0.432}^{+0.541}$ & $1.78\pm 0.04$ & $0.62$ \\
$3.721_{-0.112}^{+0.090}$ & $1.293_{-0.290}^{+0.114}$ & $1.81\pm 0.05$ & $4.16$ \\
$3.551_{-0.122}^{+0.146}$ & $1.505_{-0.227}^{+0.383}$ & $1.96\pm 0.06$ & $1.17$ \\
$3.365_{-0.142}^{+0.164}$ & $1.110_{-0.123}^{+0.081}$ & $1.83\pm 0.06$ & $7.13$ \\
$3.234_{-0.084}^{+0.185}$ & $0.589_{-0.435}^{+0.685}$ & $1.58\pm 0.04$ & $1.44$ \\
$3.076_{-0.171}^{+0.145}$ & $1.045_{-0.100}^{+0.076}$ & $1.39\pm 0.01$ & $4.50$ \\
$3.001_{-0.109}^{+0.138}$ & $0.958_{-0.142}^{+0.139}$ & $1.38\pm 0.01$ & $3.02$ \\
$2.841_{-0.129}^{+0.104}$ & $1.215_{-0.281}^{+0.145}$ & $1.411\pm 0.004$ & $1.35$ \\
$2.658_{-0.109}^{+0.221}$ & $0.899_{-0.069}^{+0.315}$ & $1.448\pm 0.004$ & $1.74$ \\
$2.593_{-0.109}^{+0.116}$ & $1.203_{-0.034}^{+0.093}$ & $1.460\pm 0.003$ & $2.75$ \\
$2.495_{-0.161}^{+0.130}$ & $1.563_{-0.110}^{+0.094}$ & $1.476\pm 0.003$ & $0.85$ \\
$2.285_{-0.058}^{+0.094}$ & $1.274_{-0.176}^{+0.301}$ & $1.503\pm 0.002$ & $0.76$ \\
$2.211_{-0.114}^{+0.117}$ & $1.310_{-0.319}^{+0.216}$ & $1.510\pm 0.002$ & $0.93$ \\
$1.961_{-0.113}^{+0.127}$ & $1.389_{-0.417}^{+0.228}$ & $1.534\pm 0.002$ & $0.64$ \\
\hline
\end{tabular}
\end{center}
\label{tbl:gamma}
\end{table}


\subsection{Results with the RSI and OCDM cosmologies}
\label{sec:cosmmodels}

We have also run our Monte-Carlo model for the RSI and the OCDM
cosmological models (see Table~\ref{tbl:cosmogony}) and have tried to
fit the model parameters to the observations of the mean optical depth
and mean temperature. We found no major differences among the results
in the three cosmological models we consider

As in the $\lcdm$ cosmology, by fitting the observations with
$\chi_T^2\le 1$ and $\chi_\tau^2\le 1$, we get $\fnq\approx 0.3$ for
both the RSI and the OCDM models. The QSOs lifetime in the RSI and the
OCDM models is $\tq\lsim 5$ Myr. A Jeans scale length of
$\xj(z\!\!=\!\!3)\approx 0.1-0.4\hmpc$ fits the RSI model, while
$\xj(z\!\!=\!\!3)\approx 0.2-0.4\hmpc$ fits the $\lcdm$, and only
$\xj(z\!\!=\!\!3)\approx 0.4\hmpc$ fits the OCDM model. While in
$\lcdm$ the bias parameter is not constrained, we find $b\approx 2-8$
and $b\approx 2-3$ for the RSI and OCDM models, respectively. As in
the $\lcdm$ cosmology, we found no constrains on the scaling function
(equation \ref{eq:scalefunc}). In the RSI and OCDM cosmological models
the initial mean temperature is $T_0(z\!\!=\!\!6)\lsim 30000$K, and
temperature increment of $\Delta T\approx 10000-15000$K. For $\Delta
T\approx 10000$K the hydrogen photo-ionization heating parameter
varies between $2.54\lsim E\lsim 3.81\ev$ while in the $\lcdm$ model
$1.27\lsim E\lsim 3.81\ev$. In all cosmological models, the heating
parameter is $1.27\ev$ for $\Delta T\approx 15000$K .


\section{Discussion}
\label{sec:discussion}

\begin{figure*}
\resizebox{0.83\textwidth}{!}{\includegraphics{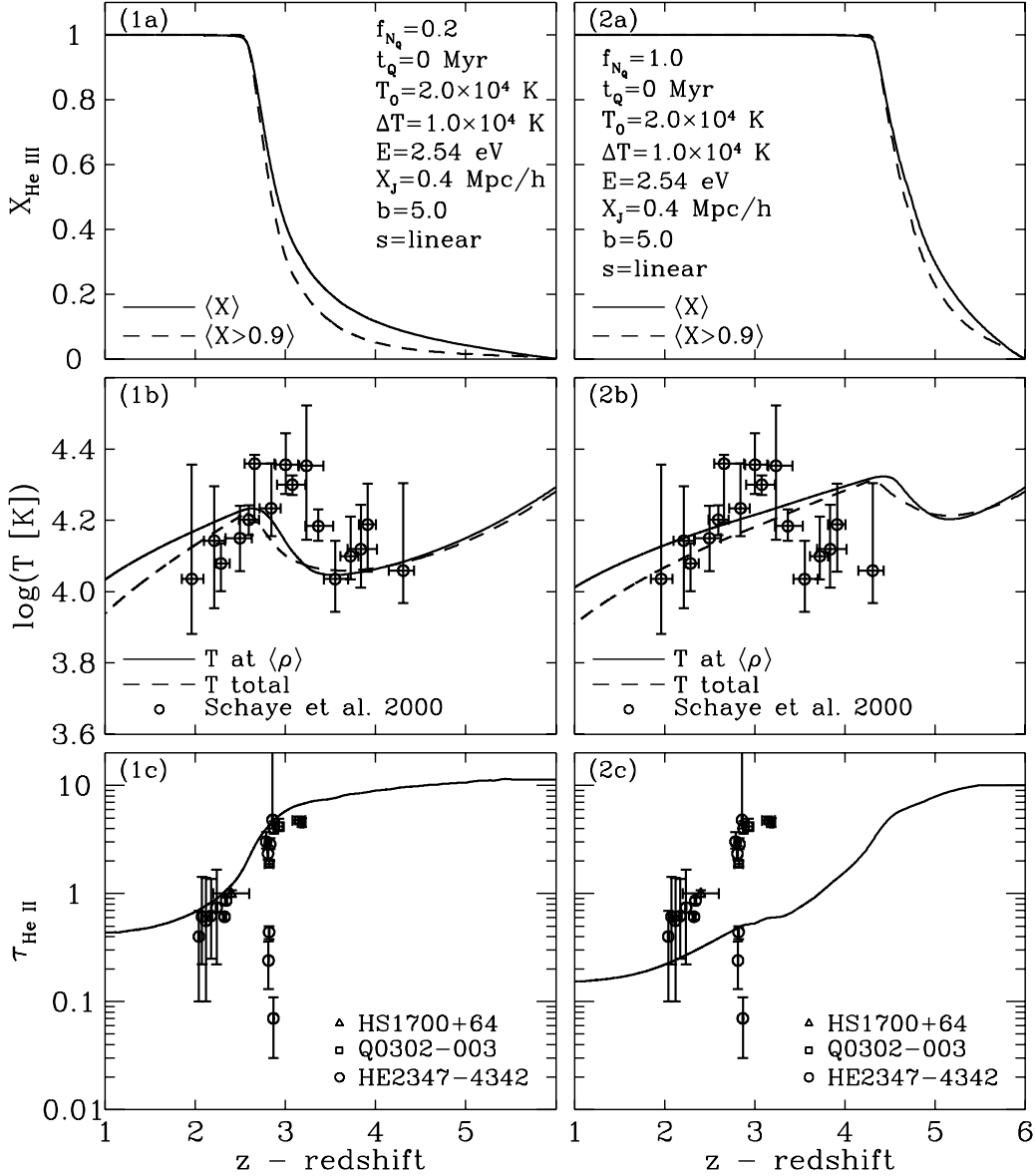}}
\caption{The filling factor, temperature and optical depth evolution
as a function of redshift in a $\lcdm$ cosmology for $\fnq=1.0$ (right
panels) $\fnq=0.2$ (left panels). For $\fnq=1.0$ the set of parameters
gives $\chi_T^2=2.3$ and $\chi_\tau^2=7.9$, and for $\fnq=0.2$,
$\chi_T^2=3.2$ and $\chi_\tau^2=0.1$. }
\label{fig:nofit}
\end{figure*}

\begin{table}
\caption{Summary of the best fit parameters derived from comparison
with the observed optical depth and temperature evolution for
$\chi_T^2\le 1$ and $\chi_\tau^2\le 1$. ``N.C.'' implies no
constraints on the parameter in the relevant range.}
\vspace{1mm}
\begin{center}
\begin{tabular}{lcccl}
\hline
 & $\lcdm$ & RSI & OCDM & \\
\hline
\hline
$f_{\nq}$ & $0.3$ & $0.3$ & $0.3$ & \\
$\tq$ & $\lsim 10$ &  $\lsim 5$ & $\lsim 5$ & Myr \\
$T_0(z\!\!=\!\!6)$ & $\lsim 4.0$ & $\lsim 3.0$ & $\lsim 3.0$ & $10^4$K \\
$\Delta T$ & $1.0-1.5$ & $1.0-1.5$ & $1.0-1.5$ & $10^4$K \\
$E$ & $1.27-3.81$ & $1.27-3.81$ & $1.27-3.81$ & $\ev$ \\
$\xj(z\!\!=\!\!3)$ & $0.2-0.4$ & $0.1-0.4$ & $0.4$ & $\hmpc$ \\
$b$ & N.C. & $2-8$ & $2-3$ & \\
$s$ & N.C. & N.C. & N.C. & \\
$z_{\rm ion}$ & $2.8-2.9$ & $2.8-2.9$ & $2.8-2.9$ & \\
\hline
\end{tabular}
\end{center}
\label{tbl:results}
\end{table}

We have presented a model for following the thermal properties of the
IGM during patchy \heii\ reionization. The model assumes that
radiation from QSOs is the main cause of \heii\ reionization and
neglects the contribution from galaxies. This is consistent with our
finding that the model can account for \heii\ reionization in a way
consistent with the available observations of the mean temperature and
\heii\ optical depth. Further, according to the {\sc galform}
semi-analytic model for galaxy formation the contribution of galaxies
to the \heii\ ionizing radiation is negligible compared to QSOs even
at redshifts as high as $z\sim 6$. We have also neglected
contributions from thermal emission in shock heated gas (Miniati \etal
2004) and also from redshifted soft X-ray produced at $z\sim 20$
(Ricotti \& Ostriker 2004). These contributions could be important for
\heii, especially at $z>3$, and will be studied in detail in a future
paper.

We have considered three different cosmological models: standard
power-law $\lcdm$, $\lcdm$ with a running spectral index (RSI), and an
open CDM and found no significant differences between them. Table
\ref{tbl:results} lists the input parameters for which the Monte-Carlo
model matches best the observational data.

A key parameter is the emission factor $\fnq$ which multiplies the
overall emission rate derived from the MHR luminosity function. In all
cosmologies, an emission factor of $\fnq=1$ yields results that are
inconsistent with the relevant observational data (see panels on the
right of figure \ref{fig:nofit}). For $\fnq=1$, reionization occurs
too early at $z=4.1$, and shifts the typical behaviour of the mean
temperature and optical depth to earlier times. Too low a value for
$\fnq$ does not match the data either. For $\fnq=0.2$ the reionization
occurs too late ($z\approx 2.5$), and the mean optical depth does not
reach high enough values around $z=3$ (see panels on the left of
figure \ref{fig:nofit}).  To match the data we need $\fnq\approx
0.3$. This result agrees with several observations (Giroux \& Shull
1997; Savaglio \etal 1997; Songaila 1998; Kepner \etal 1999; Smette
\etal 2002) and simulations (Theuns \etal 1998; Efstathiou \etal 2000)
which suggest a softer UV radiation field than the standard UV
background due to QSOs emission.

The \heiii\ fraction rises modestly at the initial stages of
reionization. The rise becomes more rapid as the \heiii\ fraction gets
close to unity (full reionization), at redshift $z_{\rm ion}$. In the
simulations of Sokasian \etal (2002) the full reionization occurs at a
redshift of $z\approx 3.8$ (see their figure 3, models 1-3), while we
found that full reionization is never achieved before $z_{\rm
ion}=3$. The reason for this difference is that we tune our model to
match the temperature evolution. In our model the temperature
increases until full reionization is reached, and declines
monotonically afterwards.

The optical depth, for \heii\ Ly$\alpha$ absorption, is
$\tau_{\nheii}\sim 10$ at $z\gsim 3.5$ with a slow decline. Around
$z_{\rm ion}$ the decline is much more rapid, from $\tau_{\nheii}\sim
10$ at $z\sim 3.5$, to $\tau_{\nheii}\sim 0.5$ at $z\sim 2.5$. At
$z\lsim 2.5$ the decline is gradual again and reached
$\tau_{\nheii}\sim 0.3$ at $z=1.0$. We need more observational data to
have better constrains on Monte-Carlo model parameters. Sokasian \etal
(2002) also compute the mean optical depth in their simulations,
however, they show results up to $z=2.8$. Some of their models also
match the observations up to that redshift, but we cannot make a more
direct comparison of their models with ours at lower redshifts.

An important application of the model is to study the
temperature-density relation in the IGM. Patchy \heii\ reionization is
expected to produce a $50\%$ scatter in this relation at $3<z<4$ (see
also Hui \& Haiman 2003). At lower redshifts, when the helium in the
IGM becomes fully ionized, photo-heating is not effective anymore and
the main process governing the gas temperature is adiabatic
cooling. The scatter is then reduced and a tight $T-\rho$ is rapidly
established. The increase in the scatter at $3<z<4$ should be taken
into account in the analysis of the Ly$\alpha$ forest data and can be
detected in high resolution spectra. The temperature-density relation
from our Monte-Carlo model fits the data from Bryan \& Machacek (2000)
around $z\sim 3$ reasonably well, but more observational data is
needed, especially in low density regions, to be able to put better
constrains on the model.

The model predicts a gradual rise in the mean temperature between
$z\sim 4$ and $z\sim 3$. Our results match the observed mean
temperature as a function of redshift (Schaye \etal 2000). A possible
discrepancy is that adiabatic cooling\footnote{Other cooling
mechanisms are inefficient at the relevant range of densities and
redshifts (Theuns \etal 1998).} is not efficient enough to cause a
rapid decline in the temperature at $z\lsim 2$, as the observations
indicate. However, the observational situation is still
inconclusive. Although the temperature increase inferred from the
observations is probably robust (Theuns \etal 2002), the quantitative
behaviour is less certain.

There is a weak degeneracy between the temperature increment, $\Delta
T$, and the energy, $E$, corresponding to photo-heating by the diffuse
hydrogen ionizing background. An energy of $E\sim 1.27\ev$ works well
with the full range $10000\lsim\Delta T\lsim 15000$K. On the other
hand the higher energies $E\sim 2.54-3.81\ev$ works only with $\Delta
T$ close to $10000$K. In any case, the inferred values for $\Delta T$
are consistent with the assumption that due to energy loses in
partially ionized regions the temperature increment must be smaller
then the estimates of Miralda-Escud\'{e} \& Rees (1994) for fully
ionized regions.

We have also found that only short QSO lifetimes fit the data:
$\tq\lsim 10$ Myr for the $\lcdm$ model and $\tq\lsim 5$ Myr for the
RSI and OCDM models. The initial mean temperature has an upper limit
of $T_0(z\!\!=\!\!6)\approx 40000$K for the $\lcdm$ model and an upper
limit of $T_0(z\!\!=\!\!6)\approx 30000$K for the RSI and OCDM models.
The Jeans scale length of $\xj(z\!\!=\!\!3)\approx 0.1-0.4\hmpc$ fits
the RSI model, while $\xj(z\!\!=\!\!3)\approx 0.2-0.4\hmpc$ fits the
$\lcdm$, and only $\xj(z\!\!=\!\!3)\approx 0.4\hmpc$ fits the OCDM
model. While in $\lcdm$ the bias parameter is not constrained, in RSI
$b\approx 2-8$ and in OCDM $b\approx 2-3$. In all cosmologies we found
no constrains on the scaling function.

In numerical hydrodynamical simulations of the Ly$\alpha$ forest
\heii\ reionization is often modelled as a result of a uniform
ionizing background of photons. This recipe for \heii\ reionization
leads to a sudden jump in the mean IGM temperature at about $z\sim
3$. Our findings imply that this recipe is oversimplified and may lead
to incorrect conclusions. An incorporation of models similar to ours
in this type of simulation is needed.

The distribution of the column densities ratio $\eta=N(\heii)/N(\hi)$
contains important information on the properties of the ionizing
sources (e.g. FUSE observations, Zheng et al. 2004). Therefore, it
will be very interesting to look at the distribution of $\eta$ as a
diagnostic of the reionization epoch. To do so one must include \hi\
reionization, and therefore to take into account the ionizing
radiation from galaxies. It is not trivial to add it to our
Monte-Carlo model since the model is based on the assumption that the
ionizing sources are short lived, which is incorrect for
galaxies. Therefore, adding the ionizing radiation from galaxies
requires a different approach which will be presented in a future
paper.


\section*{Acknowledgements}

We thank Wei Zheng for kindly providing us with the \heii\ Ly$\alpha$
optical depth data of Zheng \etal (2004), and Carlton Baugh, Shaun
Cole, Carlos Frenk and Cedric Lacey for allowing us to use results
from their {\sc galform} semi-analytic model. We also acknowledge
stimulating discussions with Joop Schaye. LG and AN acknowledge the
support of The German Israeli Foundation for the Development of
Research, the EC RTN ``Physics of the Intergalactic Medium'' and the
United States-Israeli Binational Science Foundation (grant \#
2002352). AN acknowledges support from the Royal Society's incoming
short visits programme. AJB acknowledges support from a Royal Society
University Research Fellowship. NS is supported by the JSPS grant \#
14340290.




\appendix


\section{The QSO ionizing photon emission rate}
\label{apx:ionphotons}

Our calculation of the QSO emission rate of \heii\ ionizing photons
per unit comoving volume is based on MHR for hydrogen. First we
calculate the QSO emission rate in a flat universe without a
cosmological constant ($\omg0=1$, $\omgl=0$). Later in
Appendix~\ref{apx:fcosmo} we extend this calculation to fit all
cosmologies with any desirable $\omg0$ and $\omgl$.

\subsection{QSO luminosity function}
\label{apx:lumfunc}

Following MHR we represent the quasar blue luminosity function
($\lambda_{\rm B}=4400\Ang$) as a double power-law:
\begin{equation}
\phi(L,z) = \frac{\phi_\ast/\lz}{[L_B/\lz]^{\beta_1}+[L_B/\lz]^{\beta_2}},
\label{eq:philz}
\end{equation}
where $\beta_1$ and $\beta_2$ are the power-law indices for faint and
bright quasars, respectively. The position of the break $\lz$ is
\begin{equation}
\lz = \l0(1+z)^{\alpha_s-1}\frac{\ezz(1+\exs)}{\exz+\exs},
\label{eq:lumz}
\end{equation}
where 
\begin{equation}
\l0=L_\odot\!\times\!10^{-0.4[M_\ast(0)-M_\odot]}=4.67\!\times\!10^{29}\erghzs,
\label{eq:lzero}
\end{equation}
$M_\odot=5.48$ and $L_\odot=3.44\times10^{18}\erghzs$ are the Solar
magnitude and luminosity respectively, and the rest of the luminosity
parameters can be found in Table~\ref{tbl:params}.

In Table \ref{tbl:params} $\phi_\ast$ is given for a Hubble constant
$h=0.5$. From this value we calculate $\phi_\ast(h)$ as a
function of $h$:
\begin{equation}
\phi_\ast(h)=8.9125\times\left(\frac{h}{0.5}\right)^3\times 10^{-7}\mpc3.
\label{eq:phish}
\end{equation}

\begin{table}
\caption{Parameters of the double power-law luminosity function (Pei
1995; MHR).}
\vspace{1mm}
\begin{center}
\begin{tabular}{lc}
\hline
Parameter & Value \\
\hline
\hline
$\beta1$ .............................. & $1.64$ \\
$\beta2$ .............................. & $3.52$ \\
$z_\ast$ .............................. & $1.9$ \\
$\zeta$ ................................ & $2.58$ \\
$\xi$ ................................ & $3.16$ \\
$M_\ast(0)$ ........................ & $-22.35$ \\
$\log[\phi_\ast(h_{50})/{\rm Gpc^{-3}}]$ ... & $2.95$ \\
\hline
\end{tabular}
\end{center}
\label{tbl:params}
\end{table}

\subsection{The QSO spectral energy distribution}
\label{apx:lumspec}

The luminosity spectrum of a ``typical'' QSO is assumed to have a
power-law spectral energy distribution (SED), $L(\nu) \propto
\nu^{-\alpha_s}$, with different slopes in different wavelength ranges
(MHR),
\begin{equation}
L(\nu) \propto \left\{
\begin{array}{ll}
\nu^{-0.3} & (2500\Ang<\lambda<4400\Ang) \\
\nu^{-0.8} & (1050\Ang<\lambda<2500\Ang) \\
\nu^{-1.8} & (\lambda<1050\Ang)
\end{array}\right..
\label{eq:lumspectrum}
\end{equation}
The QSO ionizing flux, $S$, is the number of ionizing photons emitted
by a QSO per second,
\begin{equation}
S=\int_{\nu_i}^{\nu_f}L(\nu)\frac{\dd\nu}{h\nu},
\label{eq:nphoton}
\end{equation}
where $h$ is the Planck constant. For convenience we represent $S$ as
a function of wavelength, $\lambda$, and since the ionization
wavelength of helium ($\lambda=228\Ang$) is less then $1050\Ang$,
\begin{equation}
S=\frac{L_B}{1.8h}\left(\frac{4400}{2500}\right)^{-0.3}\left(\frac{2500}{1050}\right)^{-0.8}\left(\frac{1050}{\lambda}\right)^{-1.8}\left|
\begin{array}{l}^{228\Ang} \\ _{0\Ang}\end{array}\right.,
\label{eq:ionflux}
\end{equation}
where $L_B$ is the luminosity in the blue band (in units of
$\erghzs$), and neglecting ionization of heavier elements.

\subsection{QSO ionizing flux function}
\label{apx:ionfluxfunc}

Equations (\ref{eq:philz}) and (\ref{eq:ionflux}) allow us to write
the number of emitted ionizing photons per unit time per unit
(comoving) volume as
\begin{equation}
\Phi(S,z) = \frac{\phi_\ast/\sz}{[S/\sz]^{\beta_1}+[S/\sz]^{\beta_2}},
\label{eq:phiflux}
\end{equation}
where $\sz$ for helium ionization is
\begin{equation}
\sz=\frac{\lz}{1.8h}\left(\frac{4400}{2500}\right)^{-0.3}\left(\frac{2500}{1050}\right)^{-0.8}\left(\frac{1050}{\lambda}\right)^{-1.8}\left|
\begin{array}{l}^{228\Ang} \\ _{0\Ang}\end{array}\right.\nonumber
\end{equation}
\begin{equation}
\ \ \ \ \ \ \ \ \ \!\!=1.057\!\times\!10^{54}(1+z)^{\alpha_s-1}\frac{\ezz(1+\exs)}{\exz+\exs}\phts.
\label{eq:szflux}
\end{equation}

\subsection{K-Correction}
\label{apx:kcorrection}

Taking into account the wavelength dependence of the power $\alpha_s$
in the QSO spectrum, one should replace the K-correction
$(1+z)^{\alpha_s-1}$ in equation (\ref{eq:szflux}), with
\begin{equation}
K_{c}(z) = \left\{
\begin{array}{ll}
(1+z)^{0.3-1} & (2500\Ang<\frac{4400\Ang}{1+z}<4400\Ang) \\
K_1 (1+z)^{0.8-1} & (1050\Ang<\frac{4400\Ang}{1+z}<2500\Ang) \\
K_1 K_2 (1+z)^{1.8-1} & (\frac{4400\Ang}{1+z}<1050\Ang)
\end{array}\right.,
\end{equation}
where $K_1=\left(\frac{4400}{2500}\right)^{0.3-0.8}$ and $K_2=\left(\frac{4400}{1050}\right)^{0.8-1.8}$.

\subsection{The QSOs emission rate of ionizing photons per unit comoving volume}
\label{apx:emisrate}

The QSO emission rate of ionizing photons per unit comoving volume can
be written as
\begin{equation}
\nq(z)=\int_{S_{min}}^{\infty}\Phi(S,z)S\dd S.
\label{eq:nqz1}
\end{equation}
where $S_{min}=0.018\s0$ corresponds to $M_B\approx-18$, and therefore
$L_{min}\approx0.018\l0$ (Cheng \etal 1985). Using equation
(\ref{eq:phiflux}) we write $\nq(z)$ as
\begin{equation}
\nq(z)=\phi_\ast(h)\sz\int_{0.018}^{\infty}\frac{x\dd x}{x^{\beta_1}+x^{\beta_2}},
\label{eq:nqz2}
\end{equation}
where $x\equiv S/\sz$. Numerical integration gives
\begin{equation}
\int_{0.018}^{\infty}\frac{x\dd x}{x^{\beta_1}+x^{\beta_2}}=2.298.
\label{eq:xintgral}
\end{equation}
Equations (\ref{eq:nqz2}), (\ref{eq:phish}) and (\ref{eq:szflux})
 yield the QSO emission rate of helium-ionizing photons
\begin{equation}
\nq(z)=2.16\!\times\!10^{48}K_{c}(z)\frac{\ezz(1+\exs)}{\exz+\exs}\left(\frac{h}{0.5}\right)^3\smpc.
\label{eq:nqzhi}
\end{equation}

\subsection{The cosmological factor}
\label{apx:fcosmo}

Untill now we assumed a flat universe cosmology with $\omg0=1$ and
$\omgl=0$. To obtain $\nq(z)$ in different cosmologies we multiply the
above result by a cosmological factor $f_{cosmo}(\omg0,\omgl)$ so that
\begin{eqnarray}
\nq(z)\Longrightarrow f_{cosmo}(\omg0,\omgl)\nq(z),
\label{eq:nqzcf}
\end{eqnarray}
where
\begin{equation}
f_{cosmo}(\omg0,\omgl)=\frac{\dd r_{\rm c}(1,0)/\dd z}{\dd r_{\rm c}(\omg0,\omgl)/\dd z}\nonumber
\end{equation}
\begin{equation}
\ \ \ \ \ \ \ \ \ \ \ \ \ \ \ \ \ \ \ \ =\sqrt{(1-a)\omg0+(a^3-a)\omgl+a},
\end{equation}
and $\dd r_{\rm c}/\dd z$ is the rate of change of comoving distance
with redshift.


\section{The number of absorbed photons per redshift interval bin}
\label{apx:Nabs}

We write the total number of ionizing photons that were absorbed at
time interval $[t,t+\Delta t]$ as
\begin{equation}
\nabs(z)=\left[\nq(z)\Delta
t+\ntrs\right]\left[1-e^{-c\Delta t/\lph(z)}\right],
\label{eq:ionphots}
\end{equation}
where $\nq(z)$ is the QSO ionizing photon emission rate per unit
comoving volume, $\Delta t$ is the time step, $\ntrs$ is the number of
photons received from previous time intervals, and the mean free path
of the \heii\ ionizing photons is
\begin{equation}
\lph(z)=\left[n_{\nheii}(1+\delta(z))(1+z)^3\sigma_{\rm ion}\right]^{-1},
\label{eq:mfp}
\end{equation}
where $n_{\nheii}$ is the mean comoving number density of \heii,
$\delta(z)$ is the local density perturbation, and $\sigma_{\rm ion}$
is the mean ion-photon cross section for \heii.

The ion-photon cross section at frequency $\nu$ is 
\begin{equation}
\sigma_{\rm ion}(\nu)=8.61\times 10^{-18}\ {\rm cm^2}\left(\frac{\nu}{\nu_{\nheii}}\right)^{-4}\frac{e^{-4\arctan\epsilon/\epsilon}}{1-e^{-2\pi/\epsilon}},
\label{eq:crosssec}
\end{equation}
where $\epsilon=[\nu/\nu_{\nheii}-1]^{-1/2}$, and
$\nu_{\nheii}=1.3\times^{16}{\rm Hz}$ is the frequency at the \heii\
ionization energy, $E=54.42\ev$ (Cen 1992).

Following Theuns \etal (1998) we approximate the Cen (1992) ion-photon cross section as having simple power-low dependence on frequency 
\begin{equation}
\sigma_{\rm ion}(\nu)=1.9\times 10^{-18}\ {\rm cm^2}\left(\frac{\nu}{\nu_{\nheii}}\right)^{-3}.
\label{eq:crosssec}
\end{equation}

\subsection{Frequency binning}
\label{apx:freq_bin}

Because of the frequency dependency of the cross section we divide the
frequency interval of ionizing photons into 10 bins, from
$\nu_{\nheii}$ to infinity, and calculate the total number of absorbed
photons in each bin. We define the luminosity as the photons energy
rate per frequency (in units of $\erghzs$). The luminosity of a
``typical'' QSO at wavelengths $\lambda<1050\Ang$ is assumed to behave
as $L(\nu)\propto\nu^{-1.8}$ (equation \ref{eq:lumspectrum}). To have
equal luminosity in each bin, we choose the frequency at the edge of
every bin $j$ to be
\begin{equation}
\nu^{j+1}=\nu_{1050}\left[\left(\frac{\nu^j}{\nu_{1050}}\right)^{-1.8}-f_{\rm
B}\left(\frac{\nu_{\nheii}}{\nu_{1050}}\right)^{-1.8}\right]^{-1/1.8},
\label{eq:freqj}
\end{equation}
where $\nu_{1050}=2.85\times10^{15}{\rm Hz}$ is the frequency at
wavelength $\lambda=1050\Ang$, and $f_{\rm B}=0.1$ is 1 over the
number of bins. For each bin we calculated the mean ion-photon cross
section
\begin{eqnarray}
\sigma_{\rm ion}^j & = & \sigma_0^{\rm ion}\frac{\int_{\nu^j}^{\nu^{j+1}}\left(\frac{\nu}{\nu_{\nheii}}\right)^{-3}\left(\frac{\nu}{\nu_{1050}}\right)^{-1.8}\dd\nu}{\int_{\nu^j}^{\nu^{j+1}}\left(\frac{\nu}{\nu_{1050}}\right)^{-1.8}\dd\nu}\nonumber \\
& = & \sigma_0^{\rm ion}(\nu_{\nheii})^3\frac{\int_{\nu^j}^{\nu^{j+1}}\nu^{-4.8}\dd\nu}{\int_{\nu^j}^{\nu^{j+1}}\nu^{-1.8}\dd\nu},
\label{eq:crosssecj}
\end{eqnarray}
and the QSO ionizing photon emission rate
\begin{equation}
\nq^j(z)=f_{\rm B}\nq(z).
\label{eq:emisratej}
\end{equation}

\subsection{The number of absorbed photons}
\label{apx:absorbed_photons}

To get more accurate results for the radiative transfer we first
calculate the number of redshifted photons that were transferred to a
lower frequency bin. Next, since we choose 10 frequency bin we solve a
set of 10 differential equations for the photon transmission,
\begin{equation}
\frac{\dd \ntrs^i}{\dd t}\!=\!\left(\!N_0(z)\!-\!n_{\rm He\ II}\!-\!\sum_{j=1}^{10}\!\ntrs^j(z)\!\right)\!c\sigma_{\rm ion}^i\ntrs^i(z),
\label{eq:ntrs}
\end{equation}
where $N_0(z)=\sum_{j=1}^{10}\nq^j(z)\Delta t\!+\!\ntrs^j(z)$ is the
total number of photons that are available for ionization and
$\ntrs^j$ is the previous number of transmitted photons in bin
$j$. The actual number of the absorbed ionizing photons is
\begin{eqnarray}
\nabs(z) = N_0(z) - \sum_{j=1}^{10}\ntrs^j(z).
\label{eq:ionphotsj}
\end{eqnarray}


\section{The gas density field}
\label{apx:gasdensity}

We assume that the gas density traces the dark matter density smoothed
over the Jeans length scale, $\xj$, determined by the balance of
pressure and gravitational forces. Under this assumption we compute
the variance of the gas density field, $\sigma^2$, from the dark
matter nonlinear (dimensionless) power spectrum, $\Delta^2_{NL}(k)$
and a smoothing window function, $f_k$,
\begin{equation}
\sigma^2=\int\Delta^2_{NL}(k)|f_k|^2\frac{\dd k}{k}\; .
\label{eq:gasvariance}
\end{equation}
We use the recipe outlined in Peacock (1999) to derive the nonlinear
power spectrum from the linear one. For the smoothing window we choose
a Gaussian function $f_k=e^{-k^2\xj^2/2}$ (Hui \& Gnedin 1997, Nusser 2000), where $\xj$ is the Jeans
length scale.

The clumping factor can be defined as
\begin{equation}
\fclmp=\frac{\langle\rho^2\rangle}{\langle\rho\rangle^2}=\langle(1+\delta)^2\rangle=1+\langle\delta^2\rangle,
\label{eq:fclump}
\end{equation}
where $\rho=\langle\rho\rangle(1+\delta)$ is the gas density at any
point in the universe, $\langle \rho\rangle$ is the mean gas density,
and $\delta$ is the density fluctuation.


\end{document}